\documentclass[a4paper,fleqn,usenatbib]{mnras}

\usepackage[T1]{fontenc}
\usepackage{ae,aecompl}
\usepackage{etoolbox}
\makeatletter
\patchcmd\@combinedblfloats{\box\@outputbox}{\unvbox\@outputbox}{}{%
   \errmessage{\noexpand\@combinedblfloats could not be patched}%
}%
 \makeatother

\usepackage{graphicx}	
\usepackage{amsmath}	
\usepackage{amssymb}	
\usepackage{rotating}
\usepackage{pdflscape,array,booktabs}
\usepackage{adjustbox}

\newcommand{\kms}{\,km\,s$^{-1}$}

\newcommand{\msun}{$\rm M_\odot$}

\newcommand{\arcpix}{$\rm arcsec\;pixel^{-1}$}

\newcommand{\sigmare}{$\sigma_{\rm e}$}
\newcommand{\sigmas}{$\sigma_{\ast}$}

\newcommand{\mbh}{$M_{\bullet}$}
\newcommand{\msigmas}{$M_{\bullet}-\sigma_{\star}$}

\newcommand{\msigmae}{$M_{\bullet}-\sigma_{\rm e}$}
\newcommand{\mMk}{$M_{\bullet}-M_{k, \rm bulge}$}

\def\Ha{H$\alpha$}

\def\oiii{[\ion{O}{iii}]$\lambda\lambda4959,5007$}
\def\hii{\ion{H}{ii}}

\title[Stellar velocity dispersions of galaxies]{A catalogue of nuclear 
stellar velocity dispersions of nearby galaxies from \Ha\ STIS spectra 
to constrain supermassive black hole masses}

\author[I. Pagotto et al.]{Ilaria Pagotto,$^{1}$\thanks{E-mail: ilaria.pagotto@phd.unipd.it}
Enrico Maria Corsini,$^{1,2}$
Marc Sarzi,$^{3,4}$
Bruno Pagani,$^{5}$ \and 
Elena Dalla Bont\`a,$^{1,2}$
Lorenzo Morelli$^{6}$ 
and Alessandro Pizzella$^{1,2}$ 
\\
\\
$^{1}$Dipartimento di  Fisica e Astronomia ``G. Galilei'', Universit\`a di Padova, vicolo
	  dell'Osservatorio 3, I-35122 Padova, Italy\\
$^{2}$INAF-Osservatorio Astronomico di Padova, vicolo dell'Osservatorio 2, 
      I-35122 Padova, Italy \\
$^{3}$Armagh Observatory and Planetarium, College Hill, Armagh BT61 9DG, UK \\
$^{4}$Centre for Astrophysics Research, University of Hertfordshire,  
      College Lane, Hatfield AL10 9AB, UK \\
$^{5}$D\'epartement Sciences de la Mati\`ere, ENS de Lyon, 46 all\'ee d'Italie,
F-69364 Lyon, France \\
$^{6}$Instituto de Astronomia y Ciencias Planetarias, Universidad de Atacama, Copiap\'o, Chile
}

\date{Accepted XXX. Received YYY; in original form ZZZ}

\pubyear{2018}

\begin{document}
\label{firstpage}
\pagerange{\pageref{firstpage}--\pageref{lastpage}}
\maketitle

\begin{abstract}
We present new measurements for the nuclear stellar velocity dispersion 
\sigmas\ within sub-arcsecond apertures for 28 nearby galaxies. 
Our data consist of Space Telescope Imaging Spectrograph (STIS) long-slit 
spectra obtained with the G750M grating centred on the \Ha\ spectral range. 
We fit the spectra using a library of single stellar population models 
and Gaussian emission lines, while constraining in most cases the 
stellar-population content from an initial fit to G430L STIS spectra. 
We illustrate how these \sigmas\ measurements can be useful for 
constraining the mass \mbh\ of supermassive black holes (SBHs) by 
concentrating on the cases of the lenticular galaxies NGC~4435 and 
NGC~4459. These are characterized by similar ground-based half-light 
radii stellar velocity dispersion \sigmare\ values but remarkably 
different \mbh\ as obtained from modelling their central ionised-gas 
kinematics, where NGC~4435 appears to host a significantly undermassive 
SBH compared to what is expected from the \msigmae\ relation.
For both galaxies, we build Jeans axisymmetric dynamical models
to match the ground-based stellar kinematics obtained with SAURON
integral-field spectrograph, including a SBH with \mbh\ value as
predicted by the \msigmae\ relation and using high-resolution HST
images taken with the Advanced Camera for Surveys to construct the 
stellar-mass model. By mimicking the HST observing conditions we 
use such reference models to make a prediction for the nuclear 
\sigmas\ value. Whereas this was found to agree with our nuclear 
\sigmas\ measurement for NGC~4459, for NGC~4435 the observed \sigmas\ 
is remarkably smaller than the predicted one, which further suggests 
that this galaxy could host an undermassive SBH.
\end{abstract}




\section{Introduction}

Nowadays, there is a large body of evidence supporting the idea that 
a supermassive black hole (SBH), with a mass \mbh\ ranging from $10^{6}$ 
to $10^{10}$ \msun, lies in the centre of most of galaxies.  
Over the last two decades, the SBH mass was found to correlate with 
several properties of their host galaxy and, in particular, of its 
spheroidal component, which corresponds to the entire galaxy in the 
case of ellipticals or to the bulge for lenticulars and spirals.
These correlations suggest that the SBHs and their host spheroids 
grew together, and since the slope and scatter of such correlations 
are thought to relate to the details of the SBH-galaxy coevolution 
there is constant demand for collecting \mbh\ measurements across 
a large number of galaxies encompassing a wide range of host-galaxy 
masses and morphologies \citep[see][for a review]{Kormendy2013}.

The relationship between \mbh\ and the stellar velocity
dispersion \sigmas\ of the spheroidal component is currently the
tightest correlation discovered so far \citep{Beifiori2012,
  Kormendy2013, vandenBosch2016}, although there are open issues about
its slope \citep{Saglia2016}, whether the SBHs in barred and ordinary
galaxies and/or classical and pseudobulges follow the same correlation
\citep{Graham2016c}, and the role of the over and undermassive SBHs
with respect to those tracing the correlation
(\citealt{SavorgnanGraham2015}, \citealt{Krajnovic2018}).
In fact, there is still the need of increasing the statistics of
\mbh\ in some specific physical ranges. Indeed \citet{vandenBosch2015}
and \citet{Shankar2016} showed that the available \mbh\ estimates and
their scaling relations suffer from a remarkable bias since less dense
galaxies are under represented in the SBH demography. The variety of
tracers, dynamical models, and algorithms adopted to recover the mass
distribution in the centre of galaxies and constrain \mbh\ is a
further complication when building an unbiased sample. 
One significant contribution in this direction was given by the
stringent \mbh\ limits obtained for more than one hundred galaxies 
by \citet{Sarzi2002}, \citet{Beifiori2009} and \citet{Pagotto2017}, 
using nuclear line-width measurement for the nebular emission 
observed at sub-arcsecond resolution with the Hubble Space Telescope 
(HST).

Outlier galaxies with \mbh\ far from the predictions
of scaling relations represent an interesting challenge for
understanding the coevolution of galaxies and SBHs. High
\mbh\ outliers include NGC~4468b \citep{Kormendy1997} and NGC~1277
\citep{vandenBosch2012}, which are characterized by a steep inward
\sigmas\ gradient with central values not consistent with the
Faber-Jackson relation, NGC~4342 \citep{Cretton1999}, which is the
strongest outlier in the \mMk\ correlation, NGC~1271 \citep{Walsh2015}
and MRK~1216 \citep{Walsh2017}, which have too large \sigmas\ with
respect to their $K$-band luminosity. 
The finding of low \mbh\ outliers, or even ruling out the
  presence of a SBH is a more challenging task given to the detection
  threshold imposed by the angular resolution of present
  telescopes. Yet, remarkable low \mbh\ outlier examples do exist,
  such as the radio-loud elliptical NGC~4335 \citep{VerdoesKleijn2002}
  and the barred lenticular NGC~4435 \citep{Coccato2006}, where
  \mbh\ was constrained from the central ionised-gas kinematics
  observed with HST, or the lenticular NGC~4474 and elliptical NGC~4551, 
  where stringent \mbh\ limits were derived from stellar dynamical
  models and ground-based integral-field observations assisted by
  adaptive optics \citep{Krajnovic2018}. The merging galaxies NGC~1316
  \citep{Nowak2008} and NGC~4382 \citep{Gultekin2011} have also quite
  small \mbh\ for their galaxy luminosities but in these cases it is
  not clear if this could be linked to the merger physics and to
  limitations in search techniques.
From a theoretical perspective, models of SBH evolution
\citep[e.g., ][]{Volonteri2011} predict that about 20 per cent of
galaxies with masses between $10^9$ and $10^{10}$ \msun\ should be
characterized by \mbh\ values critically below those predicted from
present scaling relations. This is explained with an ineffective
growth of the SBHs, which also could be ejected or not formed at all
from the very beginning. Therefore, assessing the fraction of
undermassive black holes in nearby galaxies is crucial to understand
how the SBHs increased their mass and settled in galactic nuclei.

In this context, this paper provides a catalogue of new
measurements of nuclear \sigmas\ that can be useful for the
purpose of constraining \mbh. The paper is organized as follows. We
describe the sample selection in Section~\ref{sec:sample}. We measure
the nuclear \sigmas\ of the sample galaxies from Space
Telescope Imaging Spectrograph (STIS) spectra in Section~\ref{sec:fit}. 
We study in more detail NGC~4435 and NGC~4459 in Section~\ref{sec:twogal} 
to illustrate how combining Advanced Camera for Surveys (ACS) photometry 
along with Spectrographic Areal Unit for Research on Optical Nebulae 
(SAURON) integral-field spectroscopy can constrain \mbh . Finally, we 
present our conclusions in Section~\ref{sec:conclusions}. We adopt $H_{0} = \rm
70~$\kms~Mpc$^{-1}$, $\Omega_{\rm M} = 0.3$, and $\Omega_{\Lambda}=
0.7$ as cosmological parameters all through the paper.



\begin{landscape}
\begin{table}
\begin{tiny}
\caption{Properties of the sample galaxies.}
\begin{center}
\begin{adjustbox}{width=1.35\textwidth}
\begin{tabular}{cccccccccccccccccccccc}
\hline
\hline
\\
 & & & & \multicolumn{7}{c}{G430L} & & \multicolumn{7}{c}{G750M} & & \\
 \cmidrule{5-11}
   \cmidrule{13-19}
Galaxy & Morph. & Sp. Cl. & $D$ & Prop. Id. & {\emph PA} & Exp. Time & SP. Range & Bin. & \multicolumn{2}{c}{Apert.} & & Prop. Id. & {\emph PA} & Exp. Time & SP. Range & Bin. & \multicolumn{2}{c}{Apert.} & \sigmas$_{\rm ,fix}$ & \sigmas$_{\rm ,free}$ \\
 & & & [Mpc] & & [$^{\circ}$] & [h] & [\AA] &  & [\arcsec] & [pc] & & & [$^{\circ}$] & [h] & [\AA] &  & [\arcsec] & [pc] & [\kms] & [\kms]  \\
(1) & (2) & (3) & (4) & (5) & (6) & (7) & (8) & (9) & (10) & (11) & & (12) & (13) & (14) & (15) & (16) & (17) & (18) \\
\hline \\
IC0342   & SABcd(rs)       & H      & 4.3 & - & - & - & - & - & - & - & & 8591 & 13.9 & 0.80 & $6300\,$--$\,6865$ & $1\times1$ & $0.1\times0.15$  & $2\times3$ & - & $53.9^{+7.8}_{-3.8}$     \\ [0.6mm]
NGC~2685 & (R)SB0$^{+}$pec & S2/T2: & 14.3 & 8607 & 54.4 & 0.72 & $3000\,$--$\,5700$ & $1\times2$ & $0.2\times0.30$ & $14\times21$ & & 8607 & 54.4 & 0.86 & $6300\,$--$\,6865$ & $1\times2$ & $0.2\times0.30$ & $14\times21$ & $62.7^{+8.8}_{-7.8}$   & $51.0^{+9.7}_{-8.3}$  \\ [0.6mm]
NGC~3245 & SA0$^{0}$(r):?  & T2:    & 23.2 & - & - & - & - & - & - & - & & 7403 & 2.4 & 0.75 & $6300\,$--$\,6865$ & $1\times1$ & $0.2\times0.25$ & $22\times28$ & - & $260.4^{+29.2}_{-23.5}$   \\ [0.6mm]
NGC~3368 & SABab(rs)       & L2     & 17.8 & 7361 & 69.5 & 0.44 & $3000\,$--$\,5700$ & $1\times1$ & $0.2\times0.25$ & $17\times22$ & & 7361 & 69.5 & 0.75 & $6300\,$--$\,6865$ & $1\times1$ & $0.2\times0.25$ & $17\times22$ & $93.4^{+6.4}_{-6.1}$  & $103.9^{+7.9}_{-4.7}$		 \\ [0.6mm]
NGC~3379 & E1              & L2/T2:: & 18.0 & - & - & - & - & - & - & - & & 8589 & 75.3 & 1.78 & $6300\,$--$\,6865$ & $1\times1$ & $0.2\times0.25$ & $17\times22$ & - & $264.7^{+29.6}_{-42.5}$  \\ [0.6mm]
NGC~3489 & SAB0$^{+}$(rs)  & T2/S2  & 14.6 & 7361 & 59.1 & 0.46 & $3000\,$--$\,5700$ & $1\times1$ & $0.2\times0.25$  &  $14\times18$  & & 7361 & 59.1 & 0.71 & $6300\,$--$\,6865$ & $1\times1$ & $0.2\times0.25$ & $14\times18$ & $74.1^{+3.9}_{-4.0}$ & $78.2^{+2.4}_{-3.5}$  \\ [0.6mm]
NGC~3627 & SABb(s)         & T2/S2  & 15.4 & 8607 & 80.1 & 0.65 & $3000\,$--$\,5700$ & $1\times2$ & $0.2\times0.30$   & $15\times22$  & & 8607 & 80.1 & 0.79 & $6300\,$--$\,6865$ & $1\times2$ & $0.2\times0.30$ & $15\times22$ & $114.3^{+18.5}_{-13.6}$   & -  \\ [0.6mm]
NGC~3675 & SAb(s)          & T2     & 14.4 & 8607 & 25.9 & 0.69 & $3000\,$--$\,5700$ & $1\times2$ & $0.2\times0.30$  & $14\times21$ & & 8607 & 25.9 & 0.83 & $6300\,$--$\,6865$ & $1\times2$ & $0.2\times0.30$ & $14\times21$ & $110.7^{+13.7}_{-11.1}$   & $115.5^{+8.1}_{-17.2}$  \\ [0.6mm]
NGC~3992 & SBbc(rs)        & T2:    & 17.5 & 7361 & 155.3 & 0.50 & $3000\,$--$\,5700$ & $1\times1$ & $0.2\times0.25$ & $17\times21$ & & 7361 & 155.3 & 0.82 & $6300\,$--$\,6865$ & $1\times1$ & $0.2\times0.25$ & $17\times21$ & $145.9^{+13.1}_{-13.3}$   & $146.1^{+14.3}_{-11.9}$ \\ [0.6mm]
NGC~4030 & SAbc(s)         & H*       & 26.1 & 9783 & 90.7 & 0.61 & $3000\,$--$\,5700$ & $1\times1$ & $0.2\times0.25$ & $25\times32$ & & 8228 & 42.1 & 0.24 & $6490\,$--$\,7050$ & $2\times2$  & $0.2\times0.30$ & $25\times38$ & $82.3^{+27.7}_{-26.8}$   & $87.4^{+18.1}_{-16.6}$   \\ [0.6mm]
NGC~4245 & SB0/a(r):       & H      & 16.7 & 7361 & 85.7 & 0.46 & $3000\,$--$\,5700$ & $1\times1$ & $0.2\times0.25$ & $16\times20$ & & 7361 & 85.7 & 0.75 & $6300\,$--$\,6865$ & $1\times1$ & $0.2\times0.25$ & $16\times20$ & $83.8^{+7.2}_{-7.0}$ & $75.6^{+5.3}_{-4.6}$ \\ [0.6mm]
NGC~4314 & SBa(rs)         & L2     & 17.8 & 7361 & 105.3 & 0.46 & $3000\,$--$\,5700$ & $1\times1$ & $0.2\times0.25$ & $17\times22$ & & 7361 & 105.3 & 0.75 & $6300\,$--$\,6865$ & $1\times1$ & $0.2\times0.25$ & $17\times22$ & $78.2^{+9.3}_{-7.3}$  & $69.8^{+9.8}_{-6.5}$ \\ [0.6mm]
NGC~4321 & SABbc(s)        & T2     & 27.1 & 7361 & 92.9 & 0.46 & $3000\,$--$\,5700$ & $1\times1$ & $0.2\times0.25$ & $26\times33$ & & 7361 & 92.9 & 0.74 & $6300\,$--$\,6865$ & $1\times1$ & $0.2\times0.25$ & $26\times33$ & $56.9^{+12.3}_{-12.2}$    & $61.3^{+11.5}_{-12.9}$ \\ [0.6mm]
NGC~4414 & SAc(rs)?        & T2:    & 14.2 & 8607 & 125.1 & 0.67 & $3000\,$--$\,5700$ & $1\times2$ & $0.2\times0.30$ & $14\times21$ & & 8607 & 125.1 & 0.81 & $6300\,$--$\,6865$ & $1\times2$ & $0.2\times0.30$ & $14\times21$ & $82.4^{+3.9}_{-3.3}$      & $85.2^{+3.7}_{-7.8}$  \\ [0.6mm]
NGC~4429 & SA0$^{+}$(r)    & T2     & 20.5 & 8607 & 81.1 & 0.65 & $3000\,$--$\,5700$ & $1\times2$ & $0.2\times0.30$ & $20\times30$ & & 8607 & 81.1 & 0.79 & $6300\,$--$\,6865$ & $1\times2$ & $0.2\times0.30$ & $20\times30$ & $213.2^{+22.3}_{-18.3}$   & $201.7^{+22.6}_{-21.2}$ \\[0.6mm]
NGC~4435 & SB0$^{0}$(s)    & T2/H:  & 16.0 & 7361 & 89.6 & 0.46 & $3000\,$--$\,5700$ & $1\times1$ & $0.2\times0.25$  & $20\times25$ & & 7361 & 89.6 & 0.74 & $6300\,$--$\,6865$ & $1\times1$ & $0.2\times0.25$ & $20\times25$ & $87.3^{+10.2}_{-8.8}$ & $95.0^{+10.9}_{-9.4}$  \\ [0.6mm]
NGC~4459 & SA0$^{+}$(r)    & T2:    & 21.7 & 7361 & 92.9 & 0.46 & $3000\,$--$\,5700$ & $1\times1$ & $0.2\times0.25$ & $21\times26$ & & 7361 & 92.9 & 0.74 & $6300\,$--$\,6865$ & $1\times1$ & $0.2\times0.25$ & $21\times26$ & $214.6^{+11.2}_{-12.0}$   & $200.2^{+12.5}_{-10.2}$ \\[0.6mm]
NGC~4477 & SB0(s):?        & S2     & 23.8 & 7361 & 92.8 & 0.46 & $3000\,$--$\,5700$ & $1\times1$ & $0.2\times0.25$ & $23\times29$ & & 7361 & 92.8 & 0.73 & $6300\,$--$\,6865$ & $1\times1$ & $0.2\times0.25$ & $23\times29$ & $154.4^{+15.2}_{-11.5}$   & $158.4^{+11.2}_{-18.2}$ \\[0.6mm]
NGC~4501 & SAb(rs)         & S2     & 37.2 & 7361 & 91.9 & 0.46 & $3000\,$--$\,5700$ & $1\times1$ & $0.2\times0.25$ & $36\times45$ & & 7361 & 91.9 & 0.74 & $6300\,$--$\,6865$ & $1\times1$ & $0.2\times0.25$ & $36\times45$ & $102.4^{+12.3}_{-10.5}$    & $91.5^{+13.2}_{-14.8}$ \\[0.6mm]
NGC~4548 & SBb(rs)         & L2     & 11.5 & 7361 & 73.2 & 0.46 & $3000\,$--$\,5700$ & $1\times1$ & $0.2\times0.25$ & $11\times14$ & & 7361 & 73.2 & 0.74 & $6300\,$--$\,6865$ & $1\times1$ & $0.2\times0.25$ & $11\times14$ & $79.8^{+11.5}_{-8.9}$     & $81.0^{+10.2}_{-11.5}$  \\[0.6mm]
NGC~4596 & SB0$^{+}$(r)    & L2::   & 31.7 & 7361 & 70.3 & 0.46 & $3000\,$--$\,5700$ & $1\times1$ & $0.2\times0.25$ & $31\times38$ & & 7361 & 70.3 & 0.75 & $6300\,$--$\,6865$ & $1\times1$ & $0.2\times0.25$ & $31\times38$ & $219.1^{+20.0}_{-15.5}$   & $210.3^{+19.4}_{-14.8}$ \\[0.6mm]
NGC~4698 & SAab(s)         & S2     & 19.1 & 7361 & 79.0 & 0.46 & $3000\,$--$\,5700$ & $1\times1$ & $0.2\times0.25$ & $19\times23$ & & 7361 & 79.0 & 0.74 & $6300\,$--$\,6865$ & $1\times1$ & $0.2\times0.25$ & $19\times23$ & $114.9^{+9.9}_{-9.8}$   & $109.7^{+10.3}_{-9.8}$ \\[0.6mm]
NGC~4736 & (R)SAab(r)      & L2     & 7.6 & - & - & - & - & - & - & - & & 8591 & 50.1 & 1.09 & $6300\,$--$\,6865$ & $1\times1$ & $0.1\times0.15$ & $4\times6$ & - & $109.9^{+4.8}_{-7.0}$     \\
NGC~4800 & SAb(rs)         & H      & 15.5 & 7361 & 177.5 & 0.48 & $3000\,$--$\,5700$ & $1\times1$ & $0.2\times0.25$ & $15\times19$ & & 7361 & 177.5 & 0.80 & $6300\,$--$\,6865$ & $1\times1$ & $0.2\times0.25$ & $15\times19$ & $89.5^{+7.2}_{-7.6}$      & $86.6^{+10.5}_{-7.2}$  \\ [0.6mm]
NGC~4826 & (R)SAab(rs)     & T2     & 10.0 & 8607 & 88.1 & 0.65 & $3000\,$--$\,5700$ & $1\times2$ & $0.2\times0.30$ & $10\times15$ & & 8607 & 88.1 & 0.80 & $6300\,$--$\,6865$ & $1\times2$ & $0.2\times0.30$ & $10\times15$ & $82.0^{+6.5}_{-5.9}$      & $82.7^{+6.1}_{-3.7}$  \\ [0.6mm]
NGC~5055 & SAbc(rs)        & T2     & 9.8 & 7361 & 164.5 & 0.47 & $3000\,$--$\,5700$ & $1\times1$ & $0.2\times0.25$ & $9\times12$ & & 7361 & 164.5 & 0.79 & $6300\,$--$\,6865$ & $1\times1$ & $0.2\times0.25$ & $9\times12$ & $104.3^{+4.9}_{-5.1}$ & $111.1^{+2.0}_{-1.0}$  \\ [0.6mm]
NGC~7252 & (R)SA0$^{0}$:   & H*      & 64.1 & 7435 & 38.1 & 4.25 & $3000\,$--$\,5700$ & $1\times2$ & $0.1\times0.10$ & $31\times31$ & & 8669 & 126.1 & 0.19 & $6490\,$--$\,7050$ & $2\times2$ & $0.2\times0.30$ & $62\times93$ & $74.4^{+13.0}_{-11.4}$ & $77.5^{+14.0}_{-11.5}$  \\ [0.6mm]
NGC~7331 & SAb(s)          & T2     & 7.0 & 8607 & 178.9 & 0.67 & $3000\,$--$\,5700$ & $1\times2$ & $0.2\times0.30$ & $7\times10$ & & 8607 & 178.9 & 0.81 & $6300\,$--$\,6865$ & $1\times2$ & $0.2\times0.30$ & $7\times10$ & $118.9^{+6.7}_{-6.3}$  & $115.8^{+10.5}_{-6.3}$  \\ [0.6mm]
\hline
\end{tabular}
\end{adjustbox}
\end{center}
\end{tiny}
{\em Notes.} Col.(1): galaxy name. Col.(2): morphological type from
\citet[][RC3]{deVaucouleurs1991}. Col.(3): nuclear spectral class
\citep{Ho1997}, where S = Seyfert, L = LINER, H = \hii\ nucleus, T =
transition object (LINER/\hii), and 2 = type 2. Classifications 
evaluated as uncertain or highly uncertain are 
marked with a single or double colon, respectively. Classifications 
marked with * are from NASA/IPAC Extragalactic Database (NED). Col.(4):
distance from NED. The distances are obtained as $D = V_{\rm
  3K}/H_{0}$, where $V_{\rm 3K}$ is the weighted mean recessional
velocity corrected to the reference frame of the microwave background
radiation and $H_{0} = \rm 70~km~s^{-1}~Mpc^{-1}$. For IC~0342 we
assume the distance reported in \citet{Beifiori2009} and rescaled for
$H_{0} =\rm 70~km~s^{-1}~Mpc^{-1}$ since $V_{\rm 3K}$ is not provided
by NED. Col(5): HST proposal number for the G430L spectra. Col.(6):
position angle of the slit for the G430L spectra. Col.(7): total
exposure time for the G430L spectra. Col.(8): spectral range for the
G430L spectra. Col.(9): pixel binning for the G430L spectra. Col.(10):
size of the nuclear aperture within which the fit was
performed for the G430L spectra.  Col.(11): physical size of the 
nuclear aperture within which the fit was performed for the G430L
spectra. Col.(12): HST proposal number for the G750M
spectra. Col.(13): position angle of the slit for the G750M
spectra. Col.(14): total exposure time for the G750M
spectra. Col.(15): spectral range for the G750M spectra. Col.(16):
pixel binning for the G750M spectra.  Col.(17): size of the nuclear 
aperture within which the fit was performed for the G750M spectra. 
Col.(18): physical size of the nuclear aperture within which the 
fit was performed for the G750M spectra. Col.(19): nuclear \sigmas\ 
obtained using the G430L optimal template when fitting the G750M 
spectra. Col.(20): nuclear \sigmas\ obtained without using the 
G430L optimal template when fitting the G750M spectra.
\label{tab:fit}
\end{table}
\end{landscape}


\section{Sample selection}
\label{sec:sample}

In order to measure the nuclear \sigmas\ in nearby galaxies, we
looked in the Hubble Data Archive for all the STIS medium-resolution
spectra that were obtained with the slit crossing the nucleus.

We first considered the archival spectra of galaxies obtained with the
G430M grating, except for those already obtained by \citet{Krajnovic2004} 
who already measured the stellar velocity and velocity dispersion along 
the major axis of four early-type galaxies. Unfortunately, almost all the 
other spectra were heavily contaminated by strong and broad \oiii\ emission 
lines due to the presence of an AGN which prevented us to successfully 
measure the stellar kinematics.
We then looked for archival spectra obtained with the G750M grating, 
excluding also in this case the observations of \citet{Batcheldor2013} 
who already provided stellar kinematics measurements for 36 galaxies 
in the wavelength range centred on the \ion{Ca}{ii} absorption triplet 
at about 8500 \AA.  
This initial pruning left us with the same sample of 177 galaxies
compiled by \citet{Beifiori2009} for which the G750M grating was
centred on the \Ha\ line, to which we added 9 more galaxies
observed with STIS in the same wavelength range after 2009.
Among these 186 objects we then excluded those with STIS spectra
displaying a strong AGN emission and adopted a threshold for the
signal-to-noise ratio in the stellar continuum at $S/N = 15$ per
resolution element to perform reliable \sigmas\ measurements.

This selection led to a final sample composed by 28 nearby galaxies 
($D<70$ Mpc), covering a wide range of morphological types (from E 
to Scd) and nuclear activities (Seyfert 2, LINERs, \hii\ nuclei), 
which are listed in Table~\ref{tab:fit}. Almost all of them were 
observed with the 0\farcs2-wide slit and this yielded nearly the 
same spectral resolution of the template spectra we used to measure 
the stellar kinematics (see Section~\ref{subsec:starkin} for details). 
We considered only two galaxies (IC~0342 and NGC~4736) observed with 
the 0\farcs1-wide slit, owing to the superb $S/N$ of the data. All 
other archival G750M spectra obtained with the 0\farcs1-wide slit 
have too low $S/N$. 

Finally, to help with our stellar kinematics extraction based on the 
G750M spectra we also looked in the Hubble Data Archive for similarly 
centred low-resolution G430L spectra for all our sample galaxies, since 
these spectra can provide useful constraints on the nuclear stellar 
population composition (see, e.g. \citealp{Sarzi2005} and 
Section~\ref{subsec:starkin}). G430L spectra were found for 24 objects.


\section{Nuclear stellar velocity dispersion}
\label{sec:fit}

\subsection{Long-slit spectroscopy}

We retrieved from the Hubble Data Archive the STIS spectra of the
sample galaxies obtained with the G430L and G750M gratings through
either the 0\farcs1 $\times$ 52\arcsec\ or the 0\farcs2 $\times$
52\arcsec\ slit positioned on the galaxy nucleus.  The detector was a
SITe CCD with $ 1024\times1024$ pixel of $21\times21$ $\mu$m$^2$. The
G430L spectra covered the wavelength range between about
$3000\,$--$\,5700$ \AA , whereas the G750M spectra were characterized
by a wavelength range of either $6490\,$--$\,7050$ \AA\ or
$6300\,$--$\,6865$ \AA , depending on the tilt angle of the grating.
For the G750M spectra, the reciprocal dispersion was 0.554 and 1.108
\AA\ ${\rm pixel^{-1}}$ for the 1-pixel and 2-pixel binning read-out
mode along the dispersion direction, respectively. For an extended
source, this setup yielded for the 0\farcs1-wide slit an instrumental
FWHM of 5.5 and 1.1 \AA\ with the G430L and G750M grating,
respectively and it yielded for the 0\farcs2-wide slit 10.9 and 2.2
\AA\ with the G430L and G750M grating, respectively
\citep{Riley2017}. The spatial scale was 0\farcs0507 and 0\farcs101
pixel$^{-1}$ for the 1-pixel and 2-pixel binning read-out mode along
the spatial direction, respectively. The proposal number, slit width
and position angle, pixel binning, wavelength range, and total
exposure times of the STIS spectra of the sample galaxies are reported
in Table~\ref{tab:fit}.

We reduced the spectra as done in \citet{Beifiori2009} and
\citet{Pagotto2017}. We used both IRAF{\footnote{Image Reduction and
    Analysis Facility is distributed by the National Optical Astronomy
    Observatory (NOAO), which is operated by the AURA, Inc., under
    cooperative agreement with the National Science Foundation.} }
tasks and STIS reduction pipeline \citep{Dressel2007}, which we
combined in a customized IRAF procedure running the {\tt lacos\_spec}
task \citep{Dokkum2001} to remove the cosmic rays events or hot
pixels. The reduction steps included the subtraction of the overscan,
bias and dark contributions, correction for internal flat-field,
trimming of the spectra, removal of bad pixels and cosmic rays,
wavelength and flux calibration, correction for geometrical
distortion, alignment and combination of the spectra obtained for the
same galaxy with the same setup.  Finally, we averaged the innermost
spectral rows of each resulting spectrum for the purpose of analysing
a nearly squared aperture as prescribed by \citet{Beifiori2009}.


\begin{figure*}
\begin{small}
\begin{center}
{\includegraphics[width=.49\textwidth]{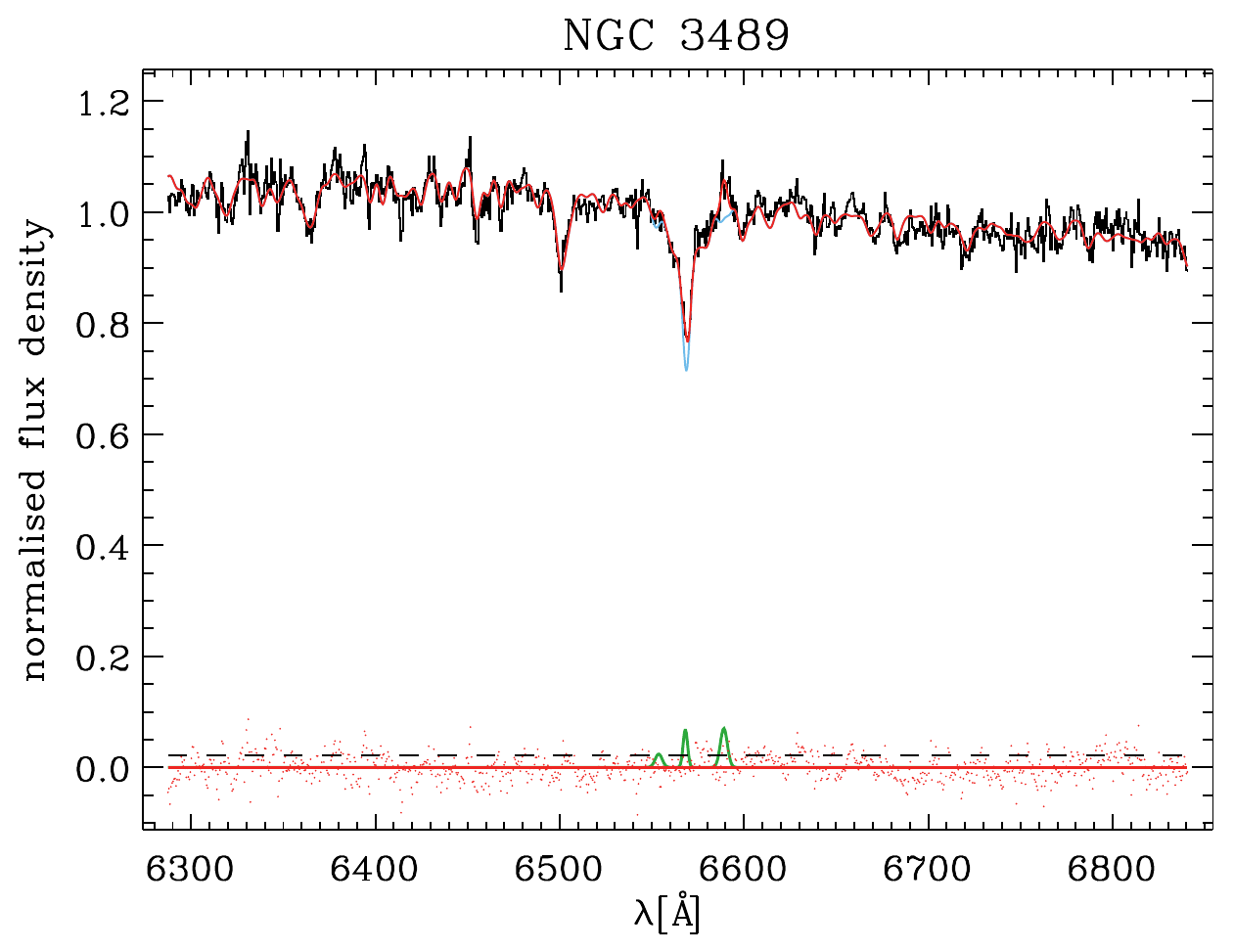}}
{\includegraphics[width=.49\textwidth]{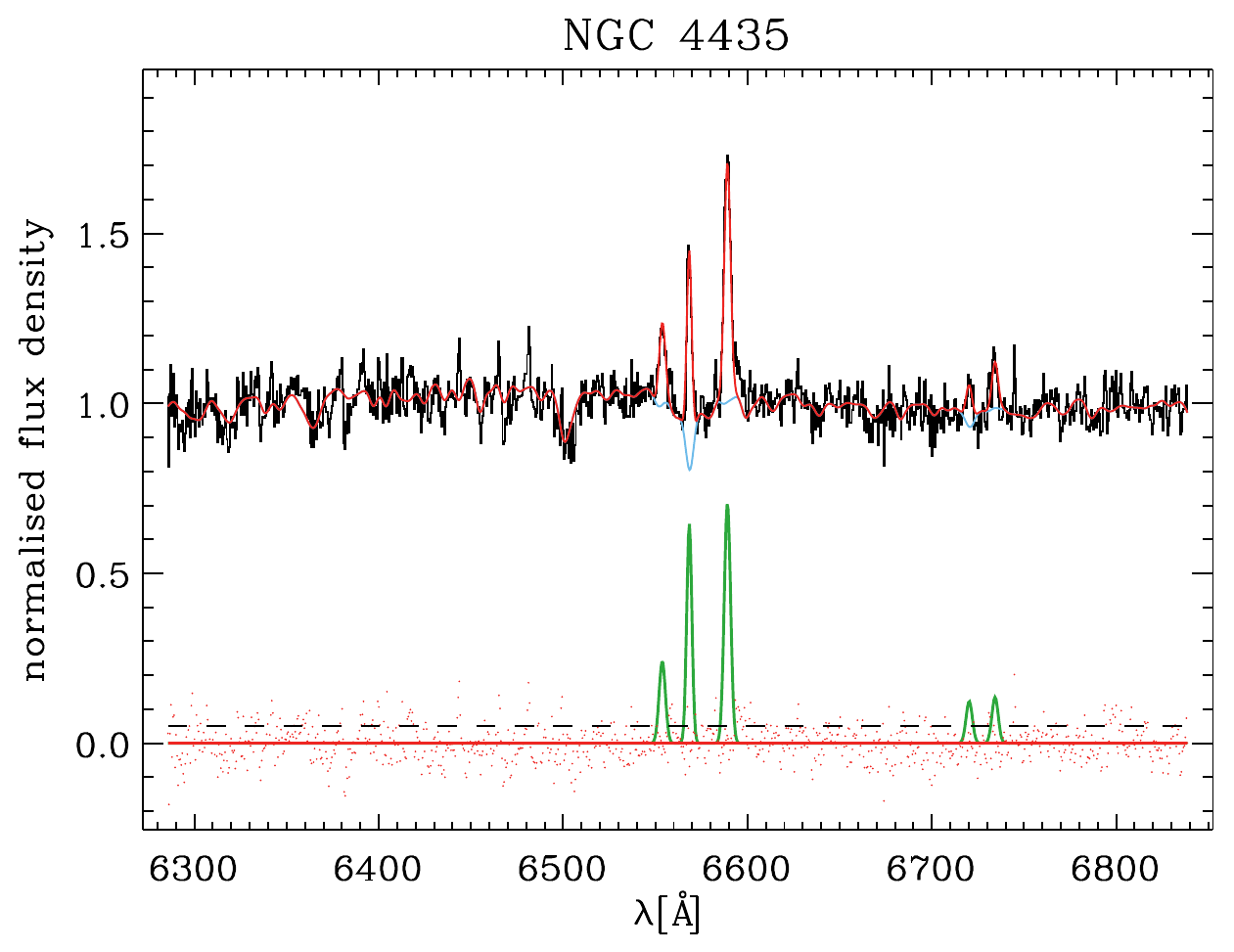}} \\
{\includegraphics[width=.49\textwidth]{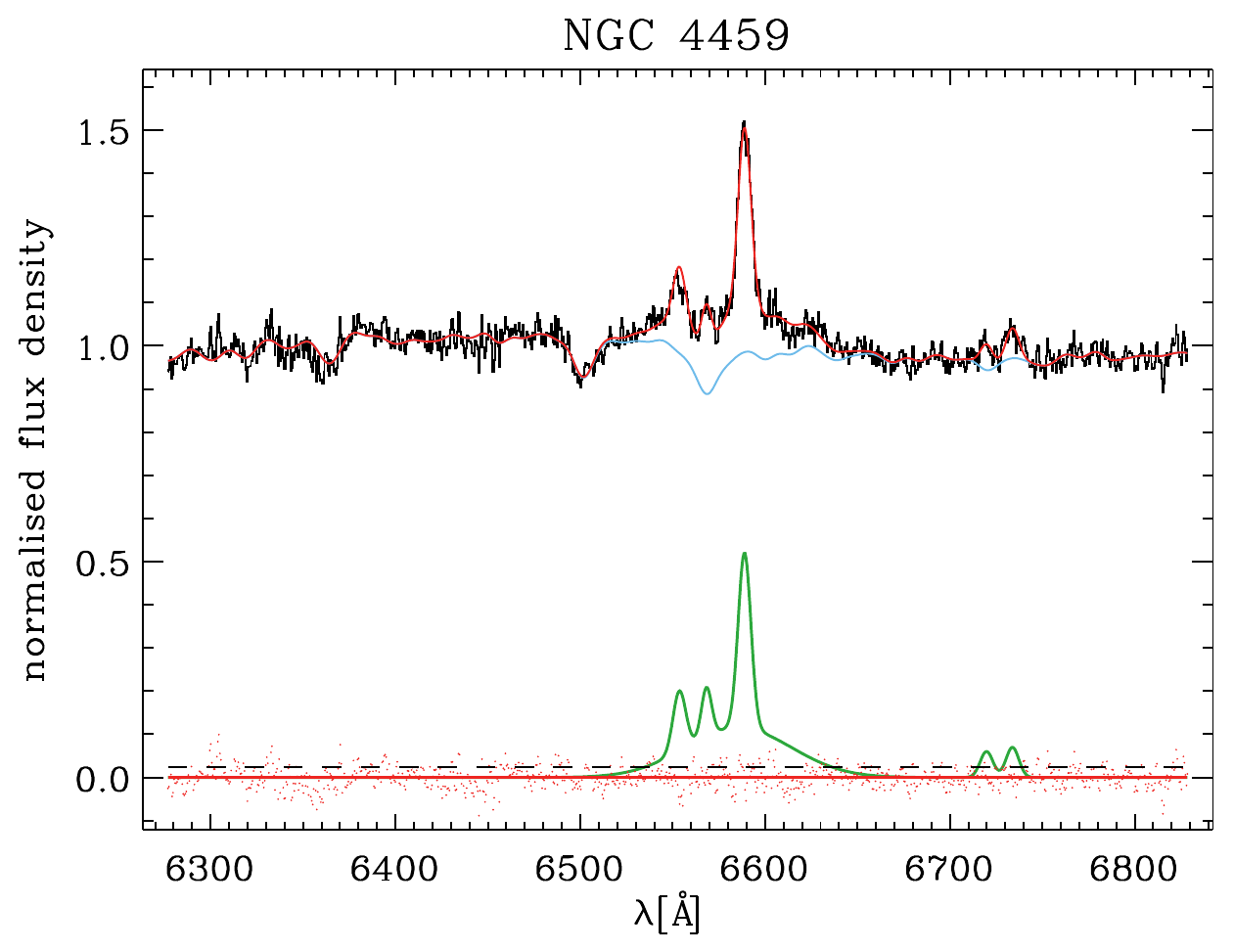}}
{\includegraphics[width=.49\textwidth]{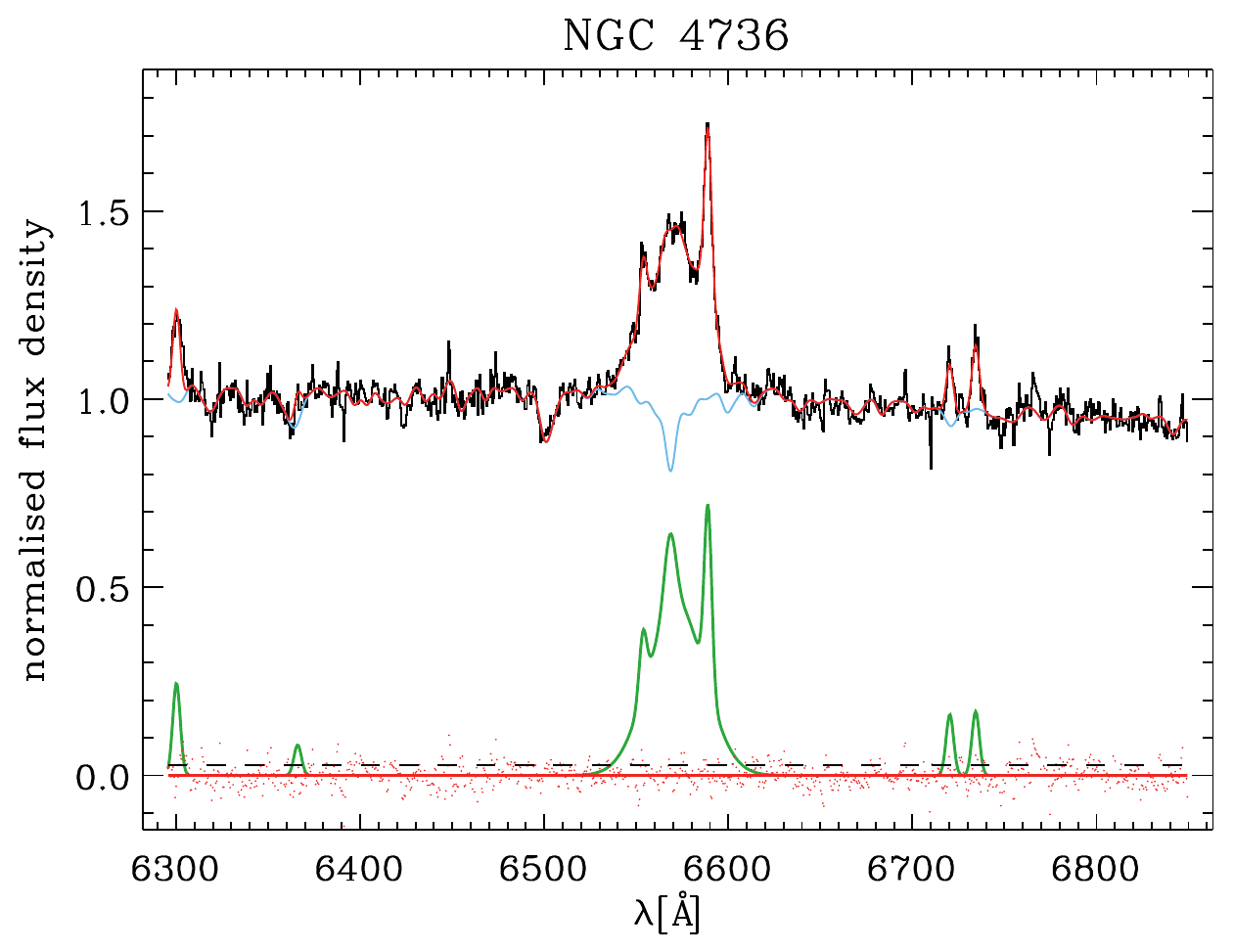}} \\
\caption{
Examples of rest-frame G750M spectra covering different cases
  with very weak narrow emission lines, bright narrow emission lines
  only, bright narrow and broad emission lines. The galaxy name is
  given in each panel and the relative fluxes have false zero points
  for viewing convenience. The best-fitting model (red line) is the
  sum of the spectra of the ionised-gas (green line) and stellar
  component (cyan line). The latter is obtained by convolving the
  synthetic templates with the best-fitting LOSVD and multiplying them
  by the best-fitting Legendre polynomials. The residuals (red dots)
  are obtained by subtracting the model from the spectrum. The dashed
  line corresponds to the rms of the residuals. 
}
\label{fig:fit}
\end{center}
\end{small}
\end{figure*}

\begin{figure*}
\begin{small}
\begin{center}
{\includegraphics[width=.239\textwidth]{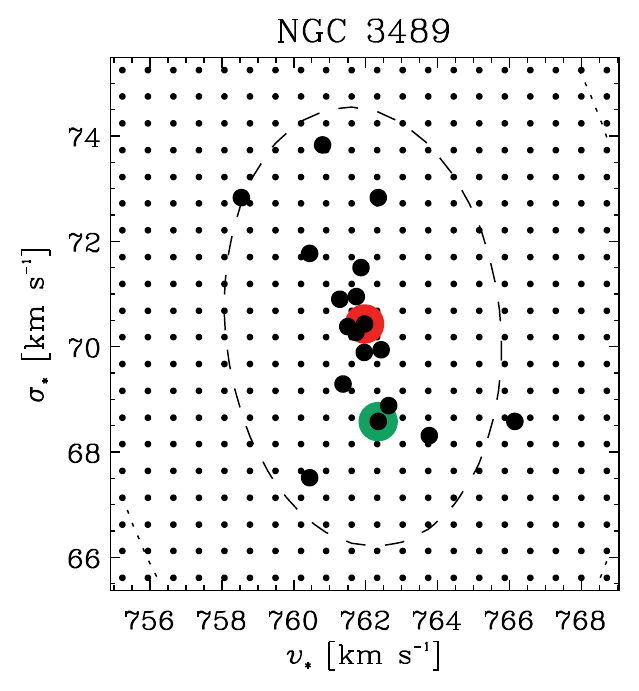}}
{\includegraphics[width=.245\textwidth]{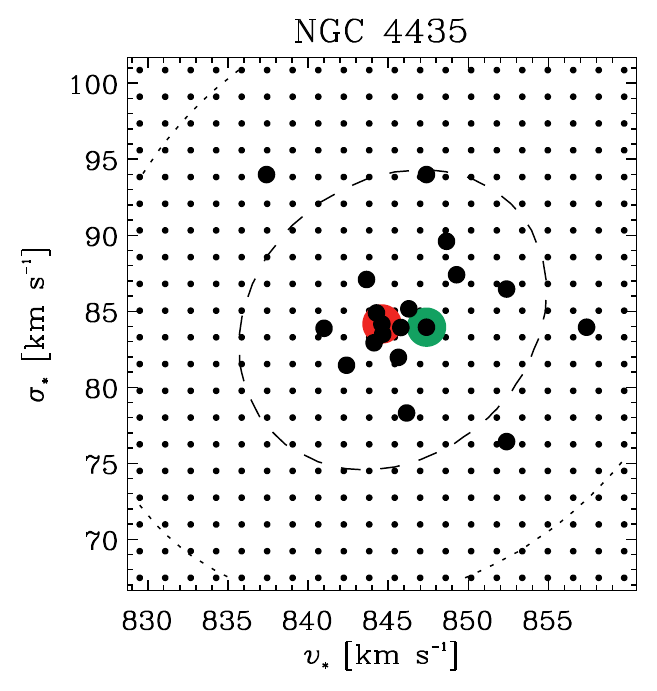}}
{\includegraphics[width=.245\textwidth]{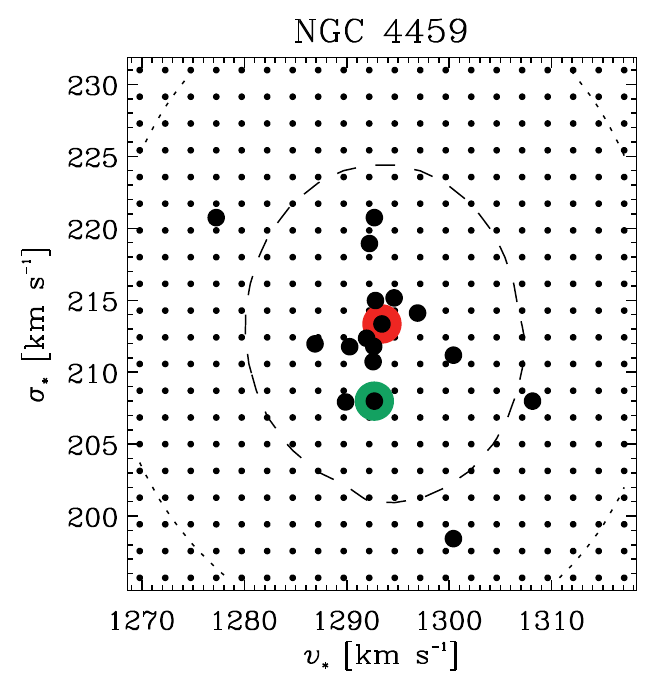}}
{\includegraphics[width=.251\textwidth]{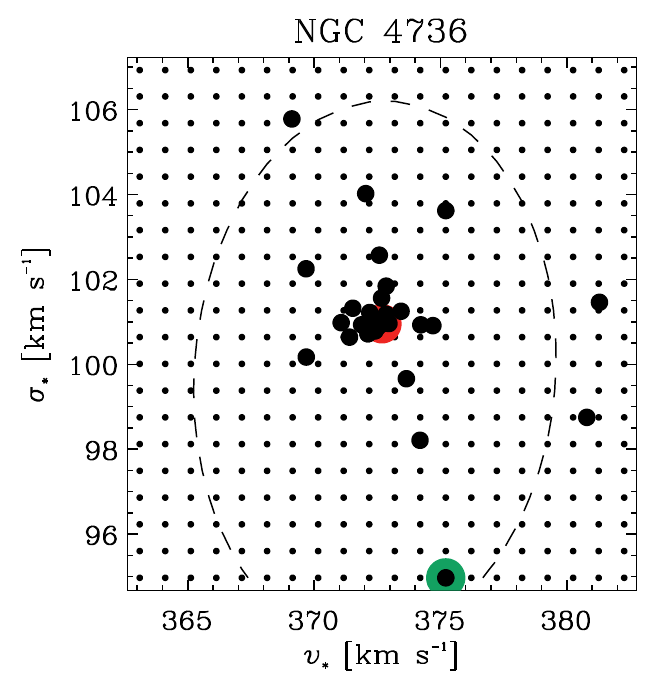}}
\caption{Grid of GANDALF models of the G750M spectra of NGC~3489,
  NGC~4435, NGC~4459, and NGC~4736 ({\em from left to right\/}) for
  different velocities and velocity dispersions of the stellar
  component (black small circles). The final best-fitting GANDALF
  model is shown with a big red circle while the starting best-fitting
  pPXF model is shown with a big green circle. Big black circles
  correspond to AMOEBA iterations. Contours show the distribution of
  $\Delta \chi^2 = \chi^2 - \chi^2_{\rm min}$ with the dashed and
  dotted lines indicating the 1$\sigma$ and 2$\sigma$ confidence level
  for two degrees of freedom, respectively.}
\label{fig:errors}
\end{center}
\end{small}
\end{figure*}

\subsection{Stellar kinematics}
\label{subsec:starkin}

\begin{figure}
	\includegraphics[scale=0.35]{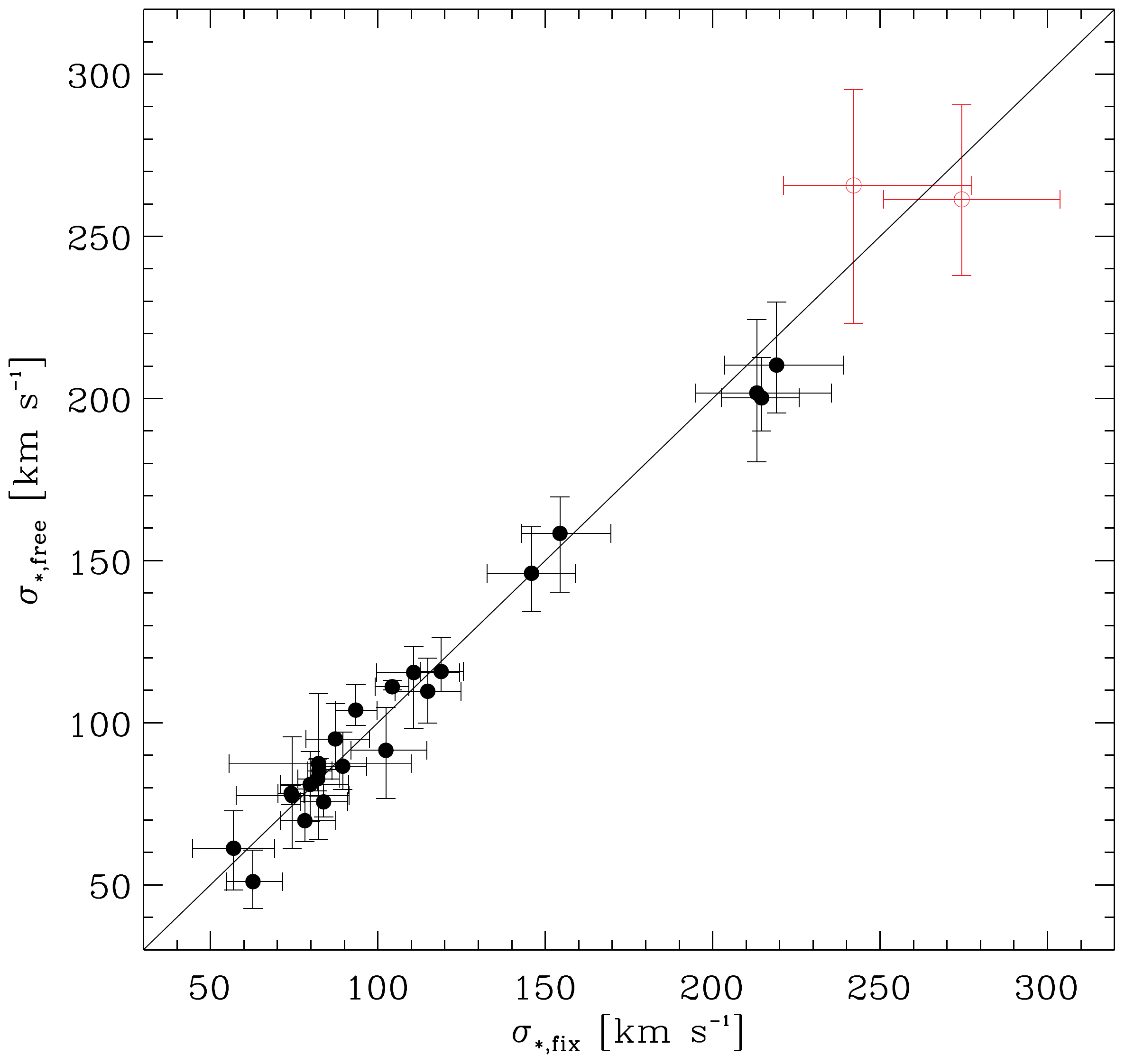}
    \caption{Comparison between the values of \sigmas$_{\rm ,fix}$ and
  \sigmas$_{\rm ,free}$ obtained from the G750M spectra of 23 sample
  galaxies (black filled circles) with and without adopting the
  optimal template from the G430L spectra, respectively. The 
  \sigmas$_{\rm ,fix}$ values of NGC~3245 and NGC~3379 
  (red open circles), for which no G430L spectrum was available, were 
  obtained by averaging the optimal templates of NGC~4429, NGC~4459, 
  and NGC~4596.}
    \label{fig:nootempl}
\end{figure}

\begin{figure}
\includegraphics[scale=0.65]{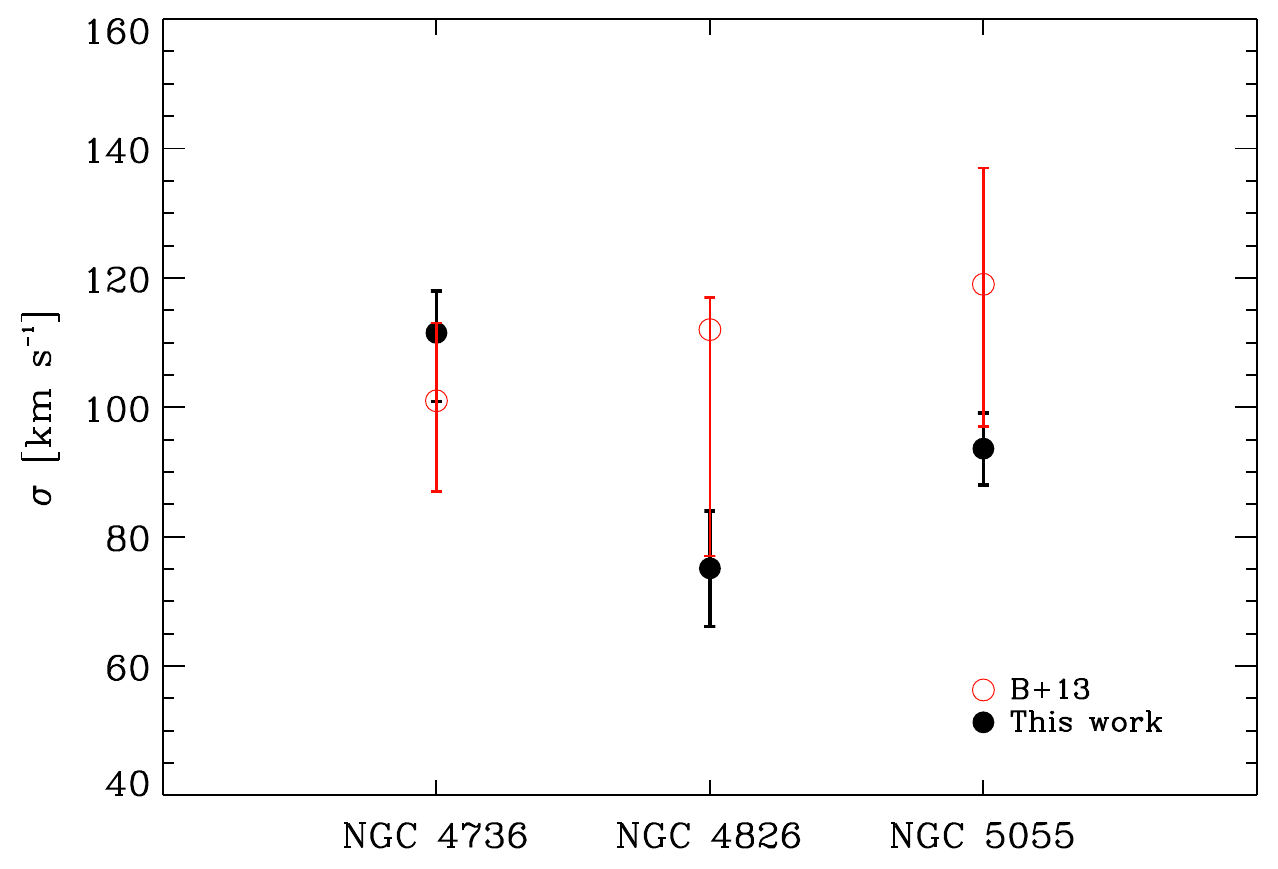}
    \caption{Comparison between our sub-arcsecond \sigmas\ measurements of
  NGC~4736, NGC~4826, and NGC~5055 from G750M spectra in the \Ha\
  region (black filled circles) and those by \citet{Batcheldor2013}
  (red open circles) from G750M spectra centred on the \ion{Ca}{ii}
  absorption triplet at about 8500 \AA .}
    \label{fig:Bath13}
\end{figure}

We measured the nuclear \sigmas\ with the Penalized Pixel Fitting 
\citep[pPXF,][]{Cappellari2004} and the Gas and Absorption Line 
Fitting \citep[GANDALF,][]{Sarzi2006} IDL\footnote{Interactive Data 
Language is distributed by Harris Geospatial Solutions.} algorithms.
In both instances, to model the stellar continuum we used the library 
of synthetic spectral energy distributions (SEDs) for single-age, 
single-metallicity stellar populations provided by \citet{Vazdekis2010}, 
which is based on the Medium Resolution Isaac Newton Telescope Library 
of Empirical Spectra \citep[MILES,][]{SanchezBlazquez2006} and covers 
the full optical spectral range of both our G430L and G750M STIS spectra 
at moderately high resolution \citep[{\it FWHM} $=2.5$ \AA ,][]{FalconBarroso2011}.

The fitting procedure consists of the following steps. We first rebinned 
each galaxy spectrum and MILES SED template along the dispersion direction 
to a common logarithmic scale. For each galaxy, we then run a preliminary 
fit to its STIS G750M spectrum using pPXF while masking the emission lines 
and considering a Gaussian LOSVD in order to obtain a starting guess on
both velocity and \sigmas. We then run a series of GANDALF fits to the 
nuclear spectrum while optimising the values for the velocity and \sigmas\ 
that GANDALF takes as input using a downhill simplex method 
\citep[AMOEBA,][]{Nelder1965}. This optimisation yields a final GANDALF 
fit and best velocity and \sigmas\ values, as well an initial estimate for 
the errors on these parameters. These error estimates are then refined by 
means of a grid of GANDALF models based on velocity and \sigmas\ values 
around the previously found best-fitting results. When necessary we also 
included a broad-line emission component in our GANDALF fits.

It should be noticed that the spectral resolution of the MILES SEDs 
is somewhat poorer than that of the G750M ones. The MILES resolution,
near the \Ha\ line corresponds to an instrumental velocity dispersion
$\sigma_{\rm inst}$ of about 49 \kms\, whereas for the G750M spectra
$\sigma_{\rm inst} = 21$ and 43 \kms\ for the 0\farcs1 and 0\farcs2-wide 
slit, respectively. During our pPXF and GANDALF fits we decided against 
degrading the resolution of our G750M spectra, since this would 
artificially smooth our spectra and lead to less reliable error 
estimates. Instead, we corrected the measured \sigmas\ to account 
for the mismatch between the instrumental resolution of our G750M data 
and the MILES templates, by summing such a difference in quadrature 
to our best-fitting \sigmas\ values. Such a correction is quite 
significant for 0\farcs1-wide slit spectra (ranging from 16 to 8
\kms\ for original \sigmas\ estimates of 50 and 100 \kms, respectively) 
whereas it remains quite small for the 0\farcs2-wide slit ones (ranging 
from 5 to less than 3 \kms\ in the same range). 

For the 24 sample galaxies with a G430L spectrum, we further re-fit
the G750M spectrum drawing from the results of a previous fit to the
G430L spectra in order to better constrain the stellar-population
content of the nuclear regions. More specifically, we performed a 
GANDALF fit to determine the weights of the best-fitting linear 
combination of MILES SEDs when modelling the G430L spectra and use 
these to construct an optimal stellar-population template that, in 
turn, can be used in our GANDALF fit to the G750M spectra. As in the 
case of the G430L the spectral resolution is significantly worse than 
that of the MILES SEDs (which is also the reason why the G430L cannot 
be used to measure \sigmas\ in first place), we proceeded to degrade 
the resolution of the latter templates to match that of G430L spectra 
before the fit, as it is common practice in these cases.

Fig.~\ref{fig:fit} and \ref{fig:errors} illustrate for four sample 
objects both the quality of our GANDALF fit to the G750M spectra as 
well as the working of our error estimation. For NGC~4736 the GANDALF 
fit employs the whole library of MILES SEDs, whereas all other examples 
correspond to fit based on the previously-derived G430L optimal template.
In this respect, Fig.~\ref{fig:nootempl} shows the comparison
between the values \sigmas$_{\rm ,fix}$ of the nuclear
velocity dispersion obtained from the G750M spectra adopting the
G430L optimal templates and values \sigmas$_{\rm ,free}$ measured
from G750M spectra using the entire MILES SED library, for all but
one of the 24 objects with both G430L and G750M spectra (for
NGC~3627 only the fit adopting the optimal template was
successful). In addition to these, in Fig.~\ref{fig:nootempl} we
also add NGC~3245 and NGC~3379 for which we adopted as G430L optimal
template the built by averaging those from the fit to NGC~4429,
NGC~4459, and NGC~4596. This is appropriate since there are good
indications that both NGC~3245 and NGC~3379 host central old stellar
population \citep{Kuntschner2010, McDermid2015} as indicated by the
stellar-population mix returned in the case of the G430L fits to
NGC~4429, NGC~4459, and NGC~4596. We found \sigmas$_{\rm ,fix}=
274.4^{+29.3}_{-23.3}$ \kms\ for NGC~3245 and \sigmas$_{\rm ,fix}=
242.1^{+35.3}_{-21.0}$ \kms\ for NGC~3379. Fig.~\ref{fig:nootempl} 
shows that the values of \sigmas$_{\rm ,fix}$ and \sigmas$_{\rm ,free}$ 
are consistent each other within $1\sigma$ errors in almost all cases
(see also Table~\ref{tab:fit}), which gives us confidence in the
\sigmas$_{\rm ,free}$ obtained for those objects (IC~0342, NGC~3245,
NGC~3379, and NGC~4736) for which G430L spectra are not available.

\citet{Batcheldor2013} already measured the nuclear
\sigmas\ of NGC~4736, NGC~4826, and NGC~5055 from G750M spectra and
0\farcs1-wide slit in the wavelength range centred on the \ion{Ca}{ii}
absorption triplet at about 8500 \AA. This allowed us to make a
comparison between \sigmas\ obtained on sub-arcsecond apertures from
G750M spectra in different wavelength ranges and with a different
fitting technique. To this aim, we re-extracted the nuclear
spectra of these galaxies in apertures matching as much as possible
that of \citet[][0\farcs1$\times$0\farcs05]{Batcheldor2013} and
repeated the fit. For NGC~4736, we were able to match the aperture and
found \sigmas$_{\rm ,free}= 111.5^{+6.5}_{-10.6}$ \kms. For NGC~4826
and NGC~5055, we could analyse only a wider aperture since our G750M
spectra were taken with the 0\farcs2-wide slit. From the spectra
extracted from the smallest possible aperture (0\farcs2$\times$0\farcs1
for NGC~4826 and 0\farcs2$\times$0\farcs05 for NGC~5055),
we measured \sigmas$_{\rm ,fix}= 75.1^{+8.9}_{-9.0}$ \kms\ for
NGC~4826 and a \sigmas$_{\rm ,fix}= 93.6^{+5.5}_{-5.6}$ \kms\ for
NGC~5055. Our measurements are in agreement with those by
\citet{Batcheldor2013} within the 1$\sigma$ error bars
(Fig.~\ref{fig:Bath13}).


\section{Stellar dynamical models for NGC~4435 and NGC~4459}
\label{sec:twogal}

In this section we focused on the particular cases of NGC~4435 and
NGC~4459 since these have very similar ground-based but quite
different STIS measurements of \sigmas, which in turn may be
indicative of a significantly different \mbh\ at a given, similar
bulge mass.

Indeed, whereas the values for \sigmare\ (the stellar velocity dispersion 
measured within a circular aperture corresponding to the galaxy effective 
radius $r_{\rm e}$) obtained by \citet{Cappellari2013} using SAURON 
integral-field spectroscopy are fairly similar for these two galaxies 
(amounting to \sigmare = $152.8\pm7.6$ and $158.1\pm7.9$ \kms\ for 
NGC~4435 and NGC~4459, respectively), our STIS measurements return a 
much smaller \sigmas$_{\rm ,fix}$ value for NGC~4435 than for NGC~4459 (amounting 
to $87.3^{+10.2}_{-8.8}$ and $214.6^{+11.2}_{-12.0}$ \kms , respectively, 
see Table~\ref{tab:fit}). In fact, the nuclear \sigmas\ value for NGC~4435 
is also at odds with what expected from the aperture velocity dispersion 
correction for early-type galaxies of \citet{FalconBarroso2017}, which 
for a $r_{\rm e}$ = 23\farcs5 would lead to a \sigmas$= 203.7 \pm 10.1$
\kms. On the other hand, for NGC~4459 and a $r_{\rm e}$ = 43\farcs1 the
same aperture correction returns a \sigmas$= 217.9 \pm 10.8$ \kms\ that 
is entirely consistent with our actual STIS measurement.
Although the use of such an aperture correction down to sub-arcsecond
scales represents almost an extrapolation given that the work of
\citet{FalconBarroso2017} is based on ground-based data, the
previous comparison further suggests that NGC~4435 and NGC~4459 have
very different central stellar kinematics.
As to whether this may indicate a different black-hole mass budget, we
note that gas-dynamical \mbh\ measurements already point in this
direction. Indeed, whereas NGC~4459 shows a \mbh\ value consistent
with the \msigmas\ relation \citep{Sarzi2001}, the \mbh\ of NGC~4435
falls significantly below it \citep{Coccato2006} in spite of their
similar \sigmare\ values.

Given these indications, we decided to use the cases of NGC~4435
and NGC~4459 to explore how our STIS sub-arcsecond
\sigmas\ measurements can be used to constrain the central mass
concentration of galaxies, in particular when large integral-field
spectroscopic measurements are available to provide further
constraints on the stellar motions.
To this goal, in what follows we describe how we proceeded to
deproject the stellar surface-brightness distribution to obtain the
galaxy luminosity density with the multi Gaussian expansion (MGE)
method of \citet{Emsellem1994} and how we built a stellar dynamical
model using the Jeans axisymmetric modelling (JAM) algorithm of
\citet{Cappellari2008}, while assuming the \mbh\ value predicted by
the \msigmae\ relation and by matching the SAURON stellar kinematic
maps provided by \citet{Krajnovic2011}.

\subsection{Properties of NGC~4435}

NGC~4435 is a barred lenticular galaxy at a distance $D=16.0$ Mpc with
a central nebular activity of intermediate type between that of LINERs 
and \hii\ nuclei (Table~\ref{tab:fit}). It is a member of the Virgo 
cluster \citep{Binggeli1985} and has an absolute total corrected $B$ 
magnitude $M^{0}_{B_{\rm T}} = -19.41$ mag, as obtained from 
$B_{\rm T} = 11.74$ mag (RC3). 

Although the proximity to the highly disturbed spiral NGC~4438 and the 
presence of optical plume that appears to connect NGC~4438 to NGC~4435 
initially lead to suggest that these two galaxies are currently interacting 
\citep{Vollmer2005}, subsequent works indicate instead that NGC~4435 is 
relatively undisturbed. \citet{Kenney2008} indeed found that NGC~4438 has 
in fact collided with the nearby elliptical M86, whereas \citet{Cortese2010} 
interpreted the optical plume between NGC~4438 and NGC~4435 as due to Galactic 
cirrus.

NGC~4435 has boxy isophotes out to about 10\arcsec\ from the centre and disky 
isophotes outwards. It hosts an inclined dust disc with blue star forming regions 
in the innermost 4\arcsec\ from the centre \citep{Ferrarese2006}. 
\citet{Coccato2006} measured an upper limit of 
$M_{\bullet}<7.5 \times 10^6$ ${\rm M_\odot}$ from the modelling of resolved, 
ionised-gas kinematics based on STIS observations along various slit directions 
and pointed out how this \mbh\ value is far below the prediction of the \msigmas\ 
relation.

\subsection{Properties of NGC~4459}

NGC~4459 is an unbarred lenticular at a distance $D=21.7$ Mpc with a
central LINER/\hii\ nebular activity (Table~\ref{fig:fit}) belonging 
to the Virgo cluster \citep{Binggeli1985}. It has 
$M^{0}_{B_{\rm T}} = -20.47$ mag from $B_{\rm T} = 11.32$ mag (RC3). 
The early-type morphology of this galaxy was investigated by several 
authors by performing a photometric decomposition. \citet{Kormendy2009} 
fitted the $V$-band surface-brightness radial profile along the major 
axis of NGC~4459 with a S\'ersic law and classified it as an elliptical 
galaxy. A S\'ersic profile was also adopted by \citet{Vika2012} and
\citet{Beifiori2012} in their two-dimensional fit of the UKIDSS
$K$-band and SDSS $i$-band images of the galaxy, respectively. On the
contrary, \citet{Sani2011} and \citet{SavorgnanGraham2016} considered
an exponential disc in addition to the S\'ersic bulge in their
photometric decomposition of the Spitzer 3.6-\micron\ image.  In their
decomposition, \citet{SavorgnanGraham2016} pointed out that the disc
starts to dominate the surface brightness profile for radii larger
than 100\arcsec. The galaxy has regular elliptical isophotes although
it hosts a dust disc which is extended out to about 8\farcs5 from the
centre and characterised by clumps of star formation
\citep{Ferrarese2006}.
From the STIS kinematics of the ionised gas \citet{Sarzi2001} measured 
a $M_{\bullet} = 9.4 \times 10^7$ ${\rm M_\odot}$ by assuming an 
inclination $i=47^{\circ}$ for the gaseous disc. This inclination value 
was determined by fitting ellipses to the innermost dust lanes, 
and is consistent with what found by \citet{Cappellari2007}
based on stellar-dynamical models.


\subsection{Broad band photometry}
\label{sec:dust}

In order to build a stellar-mass model for NGC~4435 and NGC~4459
and in turn dynamical models capable of constraining \mbh\ using our
nuclear \sigmas\ measurements, we need both high-spatial resolution
and wide-field images. Therefore, for both galaxies we retrieved
Wide Field Camera (WFC) images (Prop. Id. 9401, P.I. Patrick 
C{\^o}t\'e) obtained with the F850LP filter by ACS \citep{Lucas2016}
from the Hubble Legacy Archive (HLA). The available ACS/WFC frames
were already reduced and combined with the Python tool {\tt
  DrizzlePac} \citep{DrizzlePac2012}. The final image is characterised
by a field of view of 202\arcsec $\times$ 202\arcsec\ and a plate
scale of 0\farcs049 pixel$^{-1}$. The images were oriented with north
at the top and east to the left and given in electrons per second.

We flux calibrated the F850LP images in the AB photometric system as
follows
\begin{equation}
\mu =-2.5\log\left(\frac{I}{G \cdot s^2}\right) + z_{\rm cal} - A_z
\end{equation}
where $I$ is the surface brightness of each pixel in units of
electrons per second, $G$ is the gain in electrons per ADU, $s$ is the
plate scale in \arcpix , $z_{\rm cal}$ is the calibration constant
from the header of the images and $A_z$ is the Galactic extinction by
\citet{Schlafly2011}. It is $G=1$ electrons per ADU and $z_{\rm cal} =
24.871$ mag for both galaxies whereas it is $A_z = 0.036$ and 
0.057 mag for NGC~4435 and NGC~4459, respectively. We converted the
resulting surface brightness into the luminosity surface density in
solar units using $M_{\odot, z} = 4.52$ mag for the $z$-band absolute
AB magnitude of the Sun.

We subtracted the sky level tabulated in \citet{Pavlovsky2004}.  For
each galaxy, we did a sanity check by comparing the given value with
the mean value of the sky level we determined in a large number of
$5\times5$ pixel areas of the image using the {\tt imexamine} task in
IRAF. We selected these areas in apparently empty regions far from the
galaxy, which we considered free of objects to avoid the contamination
of the light of field stars and galaxies, as well as of the target
galaxy itself. For the estimated sky value and its error we adopted
the average and rms of the mean values, respectively. The estimated
sky value of NGC~4435 is consistent with that of
\citet{Pavlovsky2004}.  On the contrary, our estimated sky level of
NGC~4459 is significantly larger than that by \citet{Pavlovsky2004},
suggesting that the light contribution of the galaxy nearly extends
out to the edge of the field of view.

\subsection{Dust masking}

\begin{figure*}
	\includegraphics[scale=0.85]{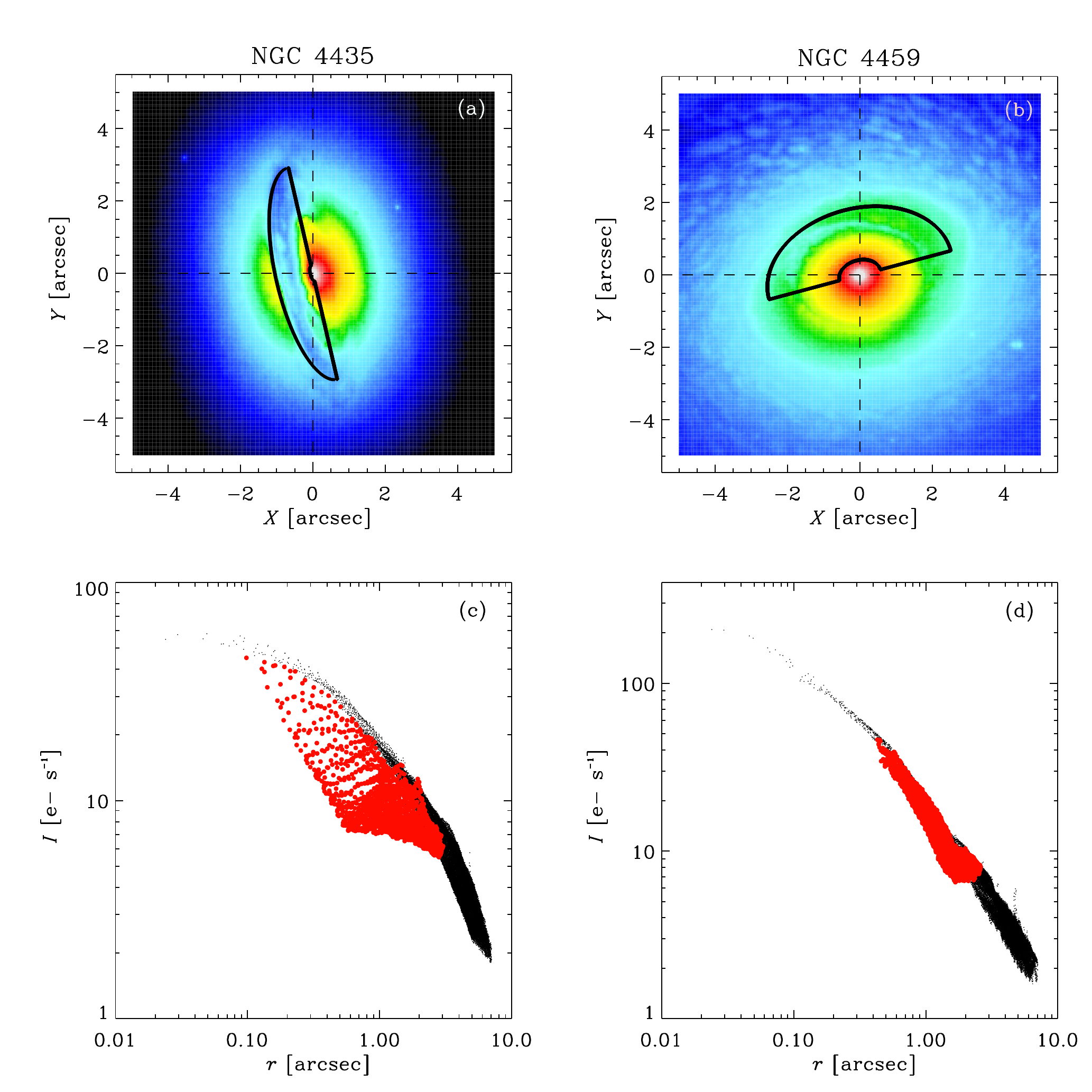}
    \caption{ {\em Left-hand panels\/}: Central portion of the F850LP
      image of NGC~4435 with the contour (black thick line) of the
      adopted mask ({\em top panel\/}) and surface brightness of the
      masked (red filled circles) and unmasked pixels (black dots) as
      a function of their distance from the galaxy centre ({\em bottom
        panel\/}). The F850LP image is oriented with north at the top
      and east to the left and the field of view is
      5\arcsec$\times$5\arcsec .  The dashed lines mark the galaxy
      centre. {\em Right-hand panels\/}: Same as above, but for
      NGC~4459.}
    \label{fig:dust}
\end{figure*}

As both galaxies display prominent dust lanes in their central
regions, we needed to account for their presence when fitting the
surface brightness distribution. First, we carefully inspected the
distribution of the dust to identify the regions where it may affect
the surface brightness analysis. Indeed, NGC~4435 harbours a
highly-inclined disc of dust near to the centre, while NGC~4459 has a
more extended dust disc with the outer parts following the surface
brightness distribution and the inner ones displaying a more irregular
pattern (Fig.~\ref{fig:dust}).

\citet{Ferrarese2006} performed a correction for dust absorption for
NGC~4435, but they failed in recovering the surface brightness profile
within the inner 2\farcs0 . The dust disc can still be seen in the
corrected image by \citet{Coccato2006} too, although it is less
optically thick than in the uncorrected image. \citet{Ferrarese2006}
considered as not reliable their dust correction of NGC~4459 out to
7\farcs5 from the centre. Therefore, we preferred to mask the dusty
regions on the images of galaxies rather than correcting them. In both
cases, the mask has the shape of half elliptical annulus with an
ellipticity $\epsilon = 1 - \cos{i}$ given by the inclination of the
dust disc. This was measured by \citet{Coccato2006} and
\citet{Ferrarese2006} for NGC~4435 ($i = 70\degr$) and NGC~4459 ($i =
45\degr$), respectively (Fig.~\ref{fig:dust}). To define the location
and size of the masked region and verify whether the galaxy centre was
obscured, we compared the surface brightness measured in each pixel of
the central portion of the image as a function of the distance from
the galaxy centre.

The mask of NGC~4435 extends between 0\farcs1 and 4\arcsec\ along the
galaxy major axis. The resulting surface brightness profile reveals a
lack of dust in the inner 0\farcs1. In fact, the unmasked pixels on
the unobscured northwestern side of the galaxy have systematically a
brighter surface brightness than those on the dusty southeastern side
(Fig.~\ref{fig:dust}).

The fact that the outer parts of the dust disc of NGC~4459 follow the
surface brightness distribution made not possible to clearly
distinguish obscured and unobscured pixels as we did in the previous
case. We chose to mask only the galaxy portion where the dust
absorption was stronger and more irregular. Therefore, we masked the
northeastern half between 0\farcs4 and 3\arcsec\ along the galaxy
major axis since there were a few dust-affected pixels in the
innermost region (Fig.~\ref{fig:dust}).


\subsection{Multi-Gaussian expansion analysis}
\label{sec:mge}

\begin{figure*}
\begin{small}
\begin{center}
{\includegraphics[width=.47\textwidth]{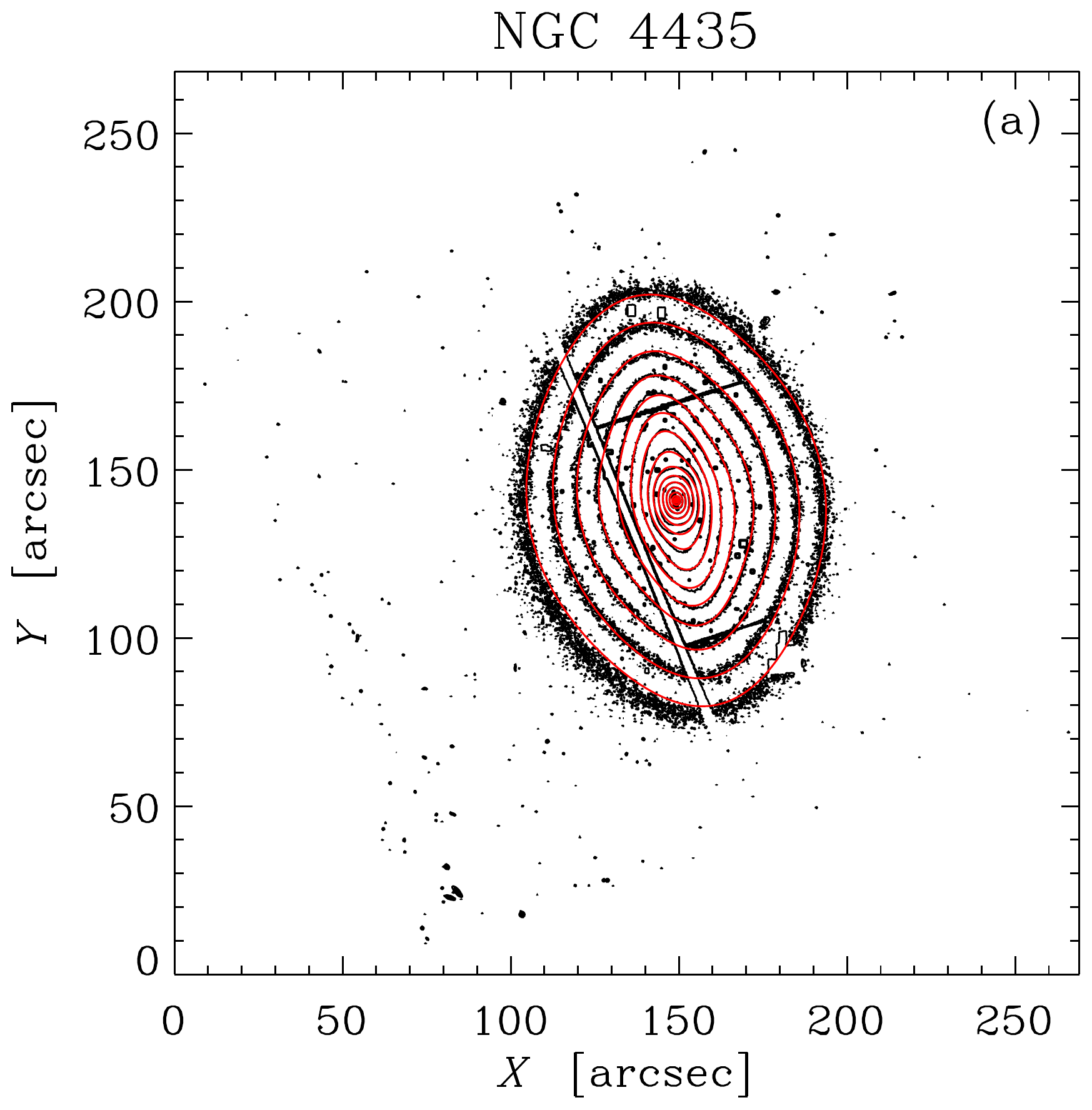}}
{\includegraphics[width=.47\textwidth]{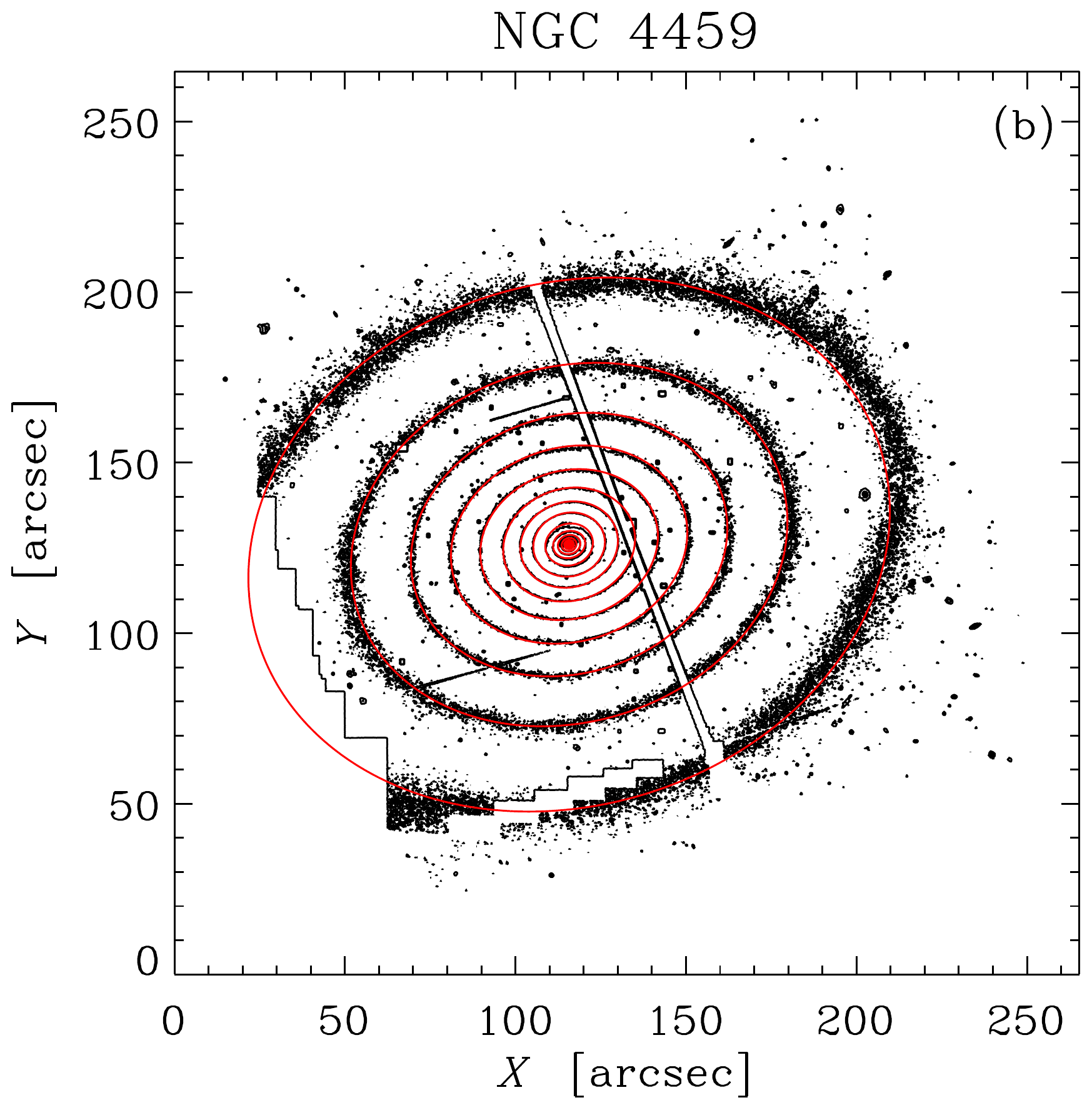}} \\
{\includegraphics[width=.47\textwidth]{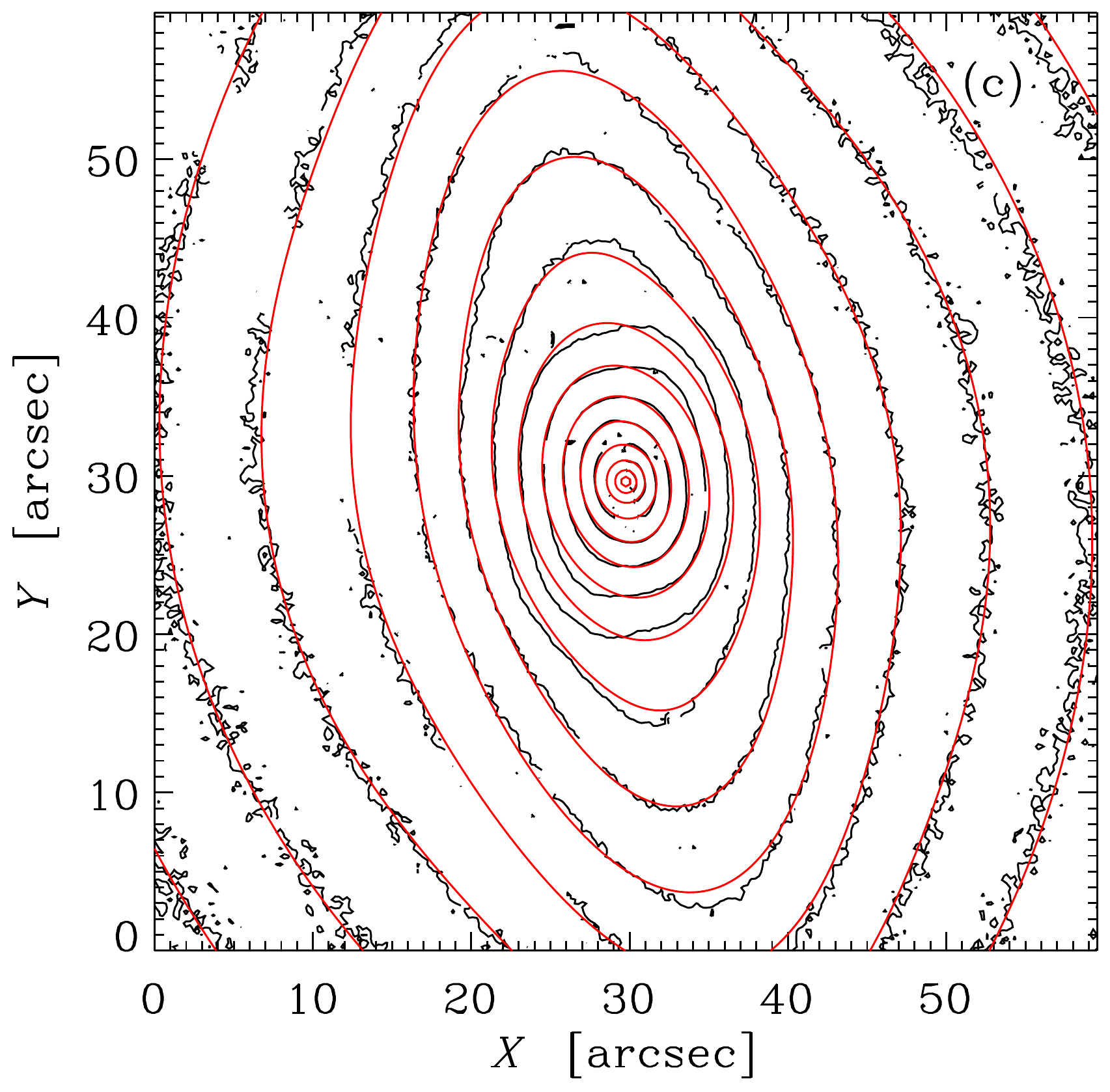}}
{\includegraphics[width=.47\textwidth]{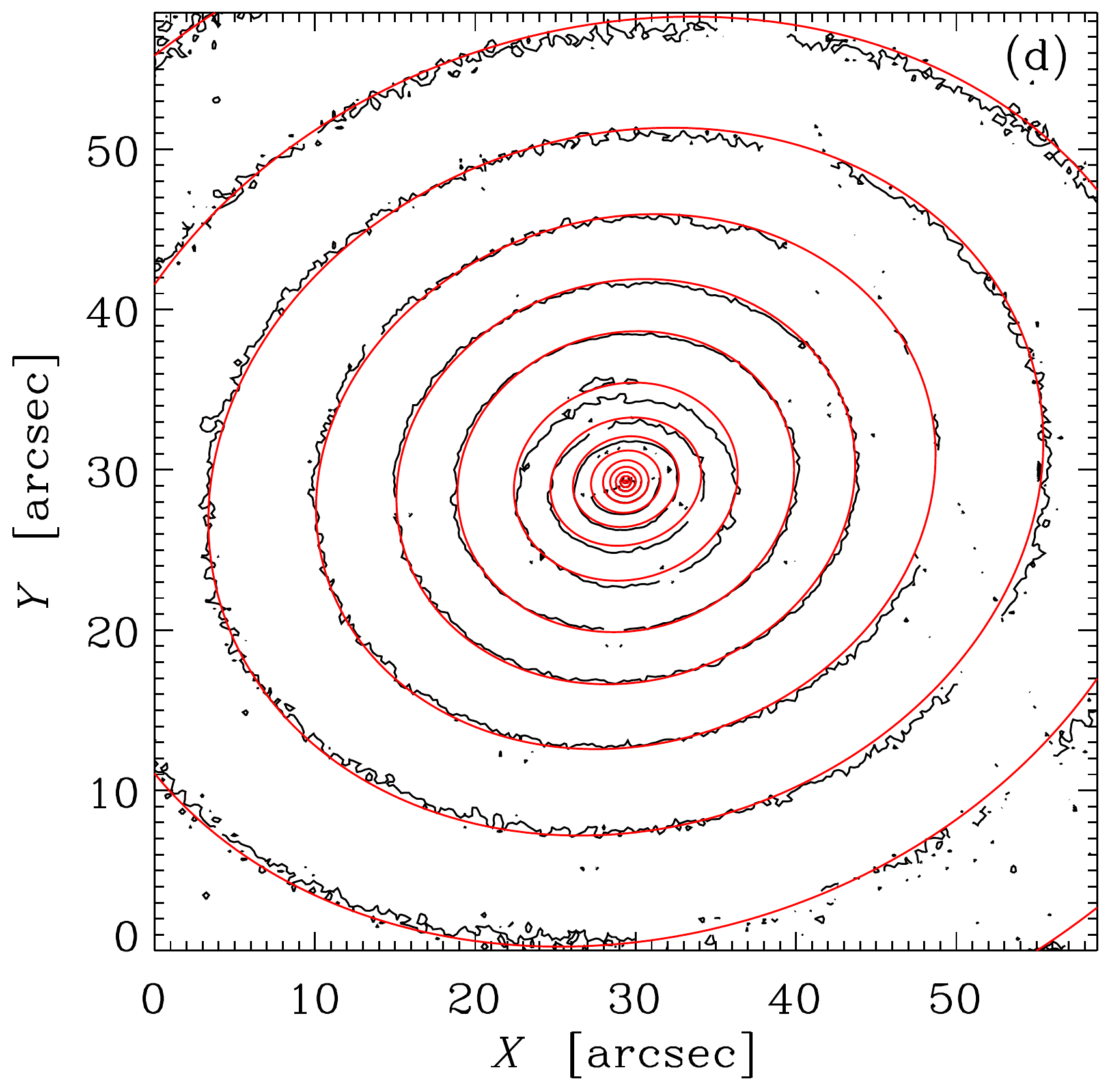}} \\
\caption{{\em Left-hand panels\/}: A few reference isophotes for the
  F850LP image (black lines) and MGE model (red lines) of NGC~4435. 
  The entire field of view ({\em top panel\/}) and the central 
  1\arcmin$\times$1\arcmin\ portion ({\em bottom panel\/}) of the 
  image are shown. The image is oriented with north at the top
  and east to the left. Flux levels are normalized to the central
  surface brightness of the image and the contours are spaced by 0.5
  mag arcsec$^{-2}$. While the MGE model was constrained using the
  original image, the image shown here is binned by $6\times6$ pixels
  to reduce the noise for comparison purposes only. The gap
  between the two WFC CCDs and a few bad rows are clearly visible,
  whereas missing portions of the galaxy isophotes correspond to masked
  regions. {\em Right-hand panels\/}: Same as above, but for
  NGC~4459. 
   }
  \label{fig:MGE}
\end{center}
\end{small}
\end{figure*}

In order to obtain a model for the luminosity volume density of
both NGC~4435 and NGC~4459, and in turn for their mass density and
gravitational potential, we started by parametrizing the surface
brightness of the sky-subtracted and dust-masked image of each
galaxy as the sum of a set of Gaussian components by using the MGE
IDL algorithm by \citet{Cappellari2002}. The MGE method indeed
allows for a simple reconstruction of the intrinsic surface
brightness distribution (provided that the PSF can be approximated as
a sum of Gaussian components) and to a straightforward deprojection of
the intrinsic surface brightness into the luminosity volume density
(which can also be expressed as the sum of a set of Gaussian components 
too).

Before using the MGE program we performed a photometric analysis
in order to estimate the centre and position angle of both galaxies. 
For this we run {\tt ellipse} task in IRAF after masking out the 
remaining foreground stars and background galaxies in the ACS images 
and inspected the results for the azimuthally-averaged surface 
brightness, ellipticity, position angle, and centre coordinates. 
In particular, we found no evidence of a varying centre within the 
errors and calculated the mean position angle by averaging the 
values measured between 3\arcsec\ and 6\arcsec\ for NGC~4435 and 
between 3\arcsec\ and 5\arcsec\ for NGC~4459.

We then obtained an MGE best-fitting model to the galaxy surface 
brightness by keeping constant the centre and position angle of the 
Gaussians and while further restricting the range of the resulting 
axial ratios of the Gaussian components but keeping the fit acceptable, 
as described in \citet{Scott2013}. This ensured that the permitted 
galaxy inclinations were not limited to a narrower range than that 
allowed by the data. We looked for the MGE best-fitting model by using 
different values of the position angle within 10\degr\ from the guess 
value we obtained from the isophotal fitting.  In the fitting process, 
we used the MGE model of the synthetic F850LP image of the PSF, which 
we generated with the TINY TIM package \citep{Krist2011}.

In Fig.~\ref{fig:MGE} we show the F850LP images of NGC~4435 and
NGC~4459 at two different scales with a few representative galaxy
isophotes and compare these to their MGE best-fitting models. 
We modeled the surface brightness radial profile of NGC~4459
as $R^{-1}$ at large radii, where \cite{SavorgnanGraham2016} 
reported the presence of an exponential disc.


\subsection{Jeans axisymmetric dynamical models}
\label{sec:jam}

With the MGE models at hand, we proceeded to use the JAM IDL algorithm 
\citep{Cappellari2008} to build Jeans axisymmetric dynamical models for 
NGC~4435 and NGC~4459 in order to test the usefulness of our nuclear 
\sigmas\ measurements in constraining the \mbh\ of these objects. 
In particular we aimed at anchoring these models by matching the JAM 
predictions for the stellar $V_{\rm rms} = \sqrt{v_{\ast}^2+\sigma_{\ast}^2}$ 
to the SAURON stellar kinematic measurements of \citet{Krajnovic2011} 
and then to obtain from such best-fitting JAM models a nuclear 
$V_{\rm rms}(0)$ value at the HST spatial resolution that can be directly 
compared to our STIS \sigmas\ measurements.

To begin with, we built a set of mass-follows-light models by assuming 
that the mass volume density follows the luminosity volume density derived 
from the MGE fit and the deprojection of the galaxy surface brightness. 
These models have three free parameters that are optimised while matching 
the observed $V_{\rm rms}$. They are the dynamical mass-to-light ratio 
$(M/L)_{\rm dyn}$, the galaxy inclination $i$ and the anisotropy parameter 
$\beta_z = 1 - \sigma_z^2 /\sigma_R^2$, where $\sigma_R$ and $\sigma_z$ 
are the radial and vertical components of the velocity dispersion, respectively, 
in a cylindrical coordinate system with the origin in the centre of the 
galaxy and symmetry axis aligned with its rotation axis.
To build such models, we took advantage of the two-dimensional maps of 
$v_{\ast}$ and \sigmas\ provided by the ATLAS$^{\rm 3D}$ survey\footnote{The 
ATLAS$^{\rm 3D}$ data are available at 
\url{http://www-astro.physics.ox.ac.uk/atlas3d/}.}  
\citep[see][for all details]{Emsellem2004, Cappellari2011, McDermid2015}. 
The integral-field spectroscopic data were obtained with SAURON working 
at William Herschel Telescope \citep{Bacon2001} in low resolution mode
with a field of view of about 30\arcsec$\times$40\arcsec . The spatial
scale per spaxel was 0\farcs8 $\times$ 0\farcs8 and the wavelength range 
between about 4800 and 5380 \AA\ was covered with spectral resolution of 
{\it FWHM} $=4.2$ \AA\ corresponding to $\sigma_{\rm inst} = 108$ \kms. 
The observations were characterised by a typical seeing of {\it FWHM} 
$=$ 1\farcs5 . The data reduction and extraction of the stellar kinematics 
was presented in \citet{Cappellari2011}, while the $v_{\ast}$ and \sigmas\ 
maps of NGC~4435 and NGC~4459 were shown in \citet[][but see also 
\citealp{Emsellem2004} for NGC~4459]{Krajnovic2011}. We derived 
$V_{\rm rms}$ and corresponding errors from the available SAURON 
kinematics with no further modification.

For NGC~4435 the MGE results allowed only for inclination angles in 
excess of $i=67\degr$ when deprojecting the surface brightness, which 
is a limit that corresponds well with inclination $i=70\degr$ for the 
central ionised-gas disc as inferred by \citet{Coccato2006} from the 
dust lane morphology. For NGC~4459 we fixed the inclination by adopting 
the value of $i=48\degr$ from \citet{Cappellari2013}, which also 
corresponds well to the $i=47\degr$ value for the central ionised-gas 
and dust disc measured by \citet{Sarzi2001}.
For both galaxies we then adopt radially constant values for both
$(M/L)_{\rm dyn}$ and $\beta_z$, which were optimised by
$\chi^2$-minimisation based on the $V_{\rm rms}$ measurements and
associated errors. For NGC~4459, considering $\beta_z<0$ would 
imply a lower $\chi^2$/DOF, but for the fast rotators the assumption 
that $\beta_z\geq0$ is observationally motivated \citep{Cappellari2008}.

The best-fitting parameters of the mass-follows-light models of
NGC~4435 were $(M/L)_{\rm dyn} = 2.18$ M$_\odot$/L$_\odot$,
$\beta_z = 0.22$ and $i=67$\degr, while for NGC~4459 we found
$(M/L)_{\rm dyn} = 1.66$ M$_\odot$/L$_\odot$ and $\beta_z = 0.00$
best-fitting values with $i=48$\degr. By construction these
mass-follows-light dynamical models do not include a SBH, so we also
constructed a second set of models including a SBH with \mbh\ as
predicted by the \msigmae\ relation of \citet{Kormendy2013} and
considering the \sigmare\ values given by \citet{Cappellari2013}, that
is, \mbh$= 9.5 \cdot 10^{7}$ and $1.1 \cdot 10^{8}$ \msun\ for NGC~4435 
and NGC~4459, respectively.
With these \mbh\ values the JAM best-fitting parameters for NGC~4435
were $(M/L)_{\rm dyn} = 2.15$ M$_\odot$/L$_\odot$, $\beta_z = 0.21$
and $i=67$\degr, while for NGC~4459 we found $(M/L)_{\rm dyn} = 1.65$
M$_\odot$/L$_\odot$ and $\beta_z = 0.00$ with
$i=48$\degr. Unsurprisingly, these are very similar to the \mbh=0
models given that ground-based data generally do not provide much
leverage on \mbh\ measurements at the distance of these galaxies.

\begin{figure*}
\begin{small}
\begin{center}
{\includegraphics[width=.50\textwidth]{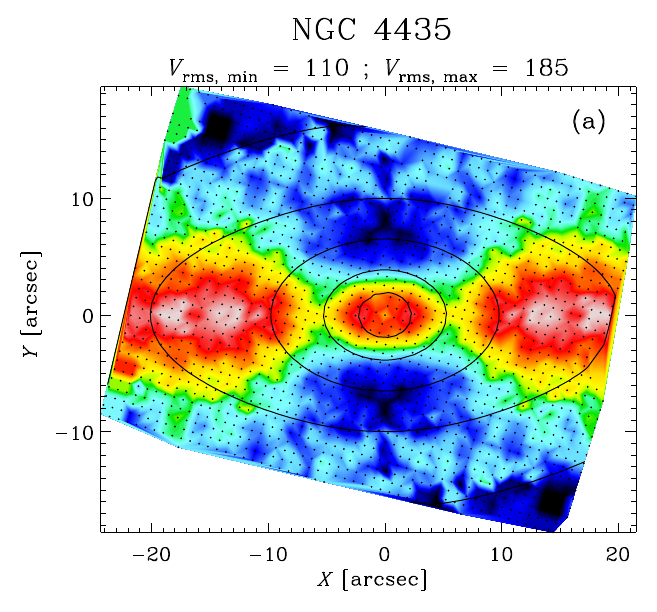}}
{\includegraphics[width=.48\textwidth]{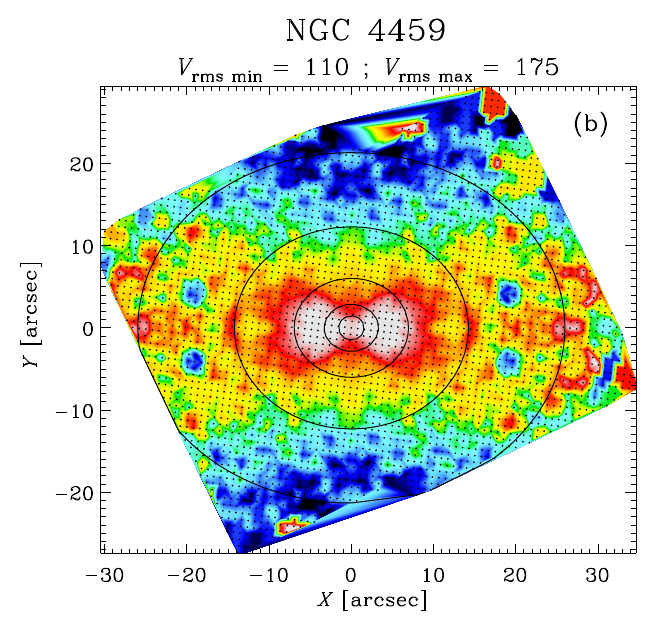}} \\
{\includegraphics[width=.50\textwidth]{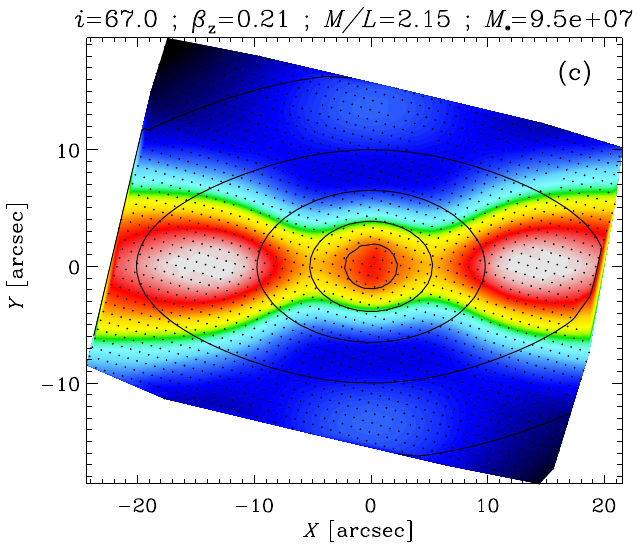}}
{\includegraphics[width=.48\textwidth]{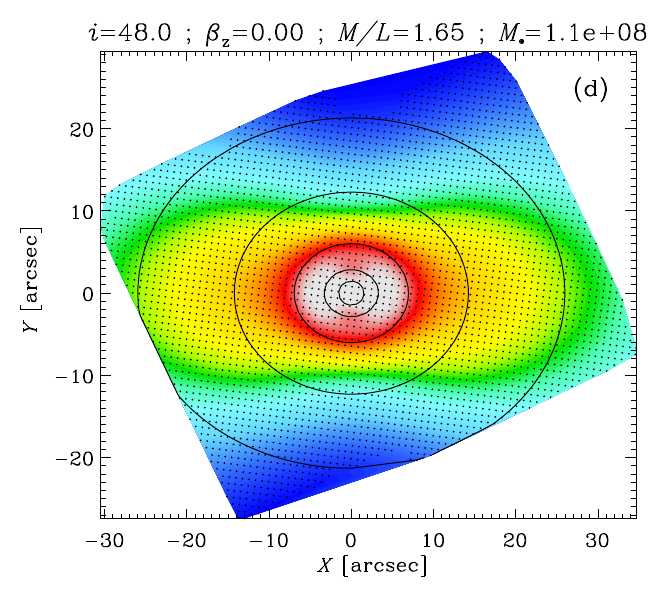}} \\
\caption{{\em Left-hand panels\/}: Comparison between $V_{\rm rms}$
  from the SAURON stellar kinematics ({\em top panel\/}) and second
  velocity moment from JAM model ({\em bottom panel\/}) of
  NGC~4435. While the models were constrained using the original
  kinematics, the symmetrized $V_{\rm rms}$ map and bisymmetric JAM
  map are shown here for comparison purposes only. The minimum and
  maximum values of $V_{\rm rms}$ corresponding to black and white
  levels in the colour table are given on top of the SAURON map. The
  best-fitting parameters of the stellar dynamical model are shown on
  top of the JAM map. A few reference isophotes for the MGE model of
  F850LP image are also plotted with the galaxy major axis parallel to
  the horizontal axis. {\em Right-hand panels\/}: Same as above, but
  for NGC~4459.  }
\label{fig:JAM}
\end{center}
\end{small}
\end{figure*}

The comparison between the SAURON $V_{\rm rms}$ maps of NGC~4435 and
NGC~4459 and JAM prediction including a SBH are shown in
Fig.~\ref{fig:JAM}. Although the models broadly compare well with the
data in these maps, formally the JAM models are not good since they
lead to reduced-$\chi^2$ values around 2.8 and 2.5, for NGC~4435 and
NGC~4459 respectively. As typical errors for the JAM best-fitting
parameters we therefore adopted the estimates that
\citet{Lablanche2012} obtained on the basis of simulations for
early-type barred and unbarred galaxies, which correspond to errors
smaller than 5$^{\circ}$ for the inclination, smaller than 0.15 for
$(M/L)_{\rm dyn}$ and smaller than 0.3 for $\beta_z$.
A comparison between our best-fitting values and those of the models
of \citet{Cappellari2013} is not possible as the latter do not provide
their $\beta_z$ values.

\begin{figure*}
\begin{small}
\begin{center}
{\includegraphics[width=.49\textwidth]{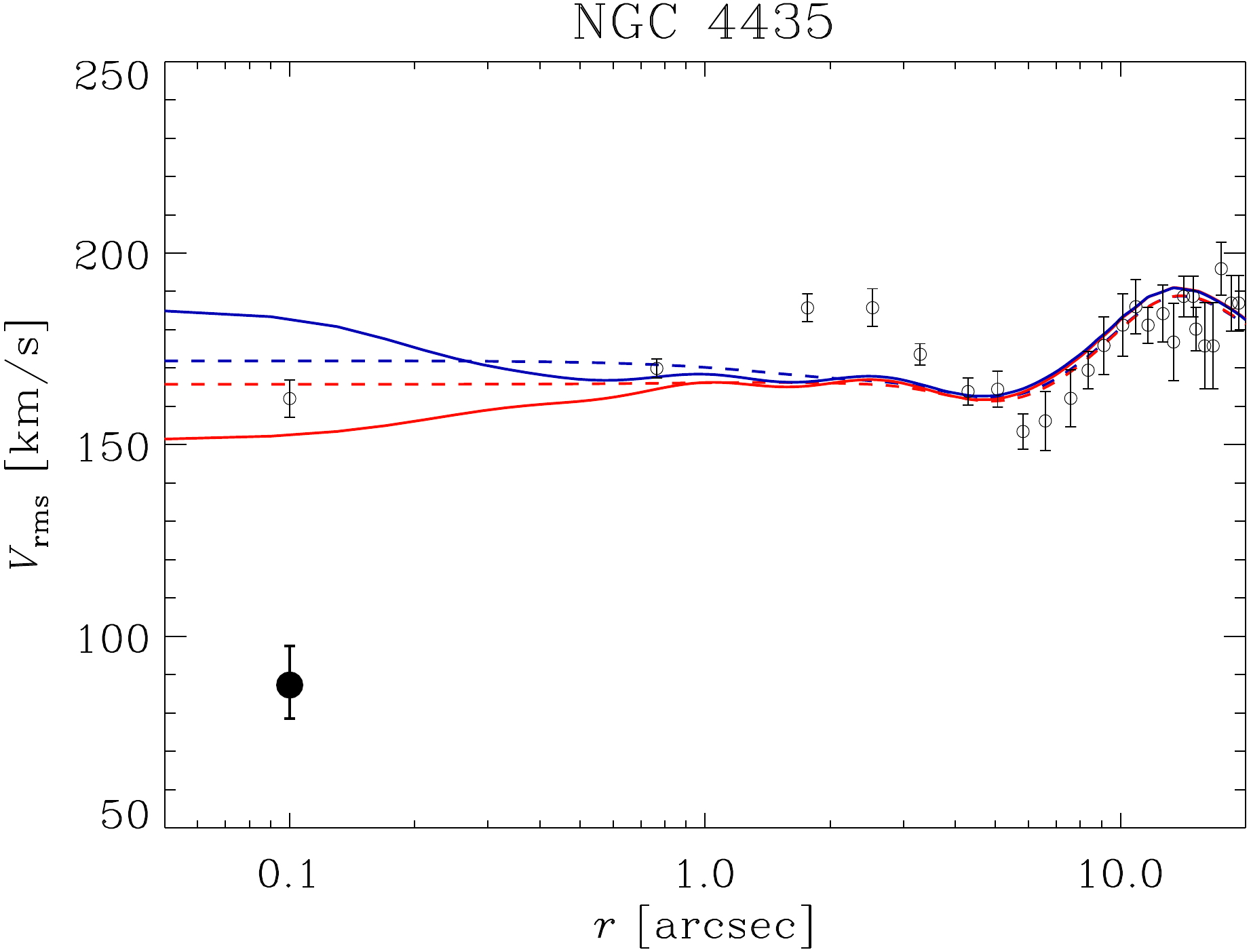}}
{\includegraphics[width=.49\textwidth]{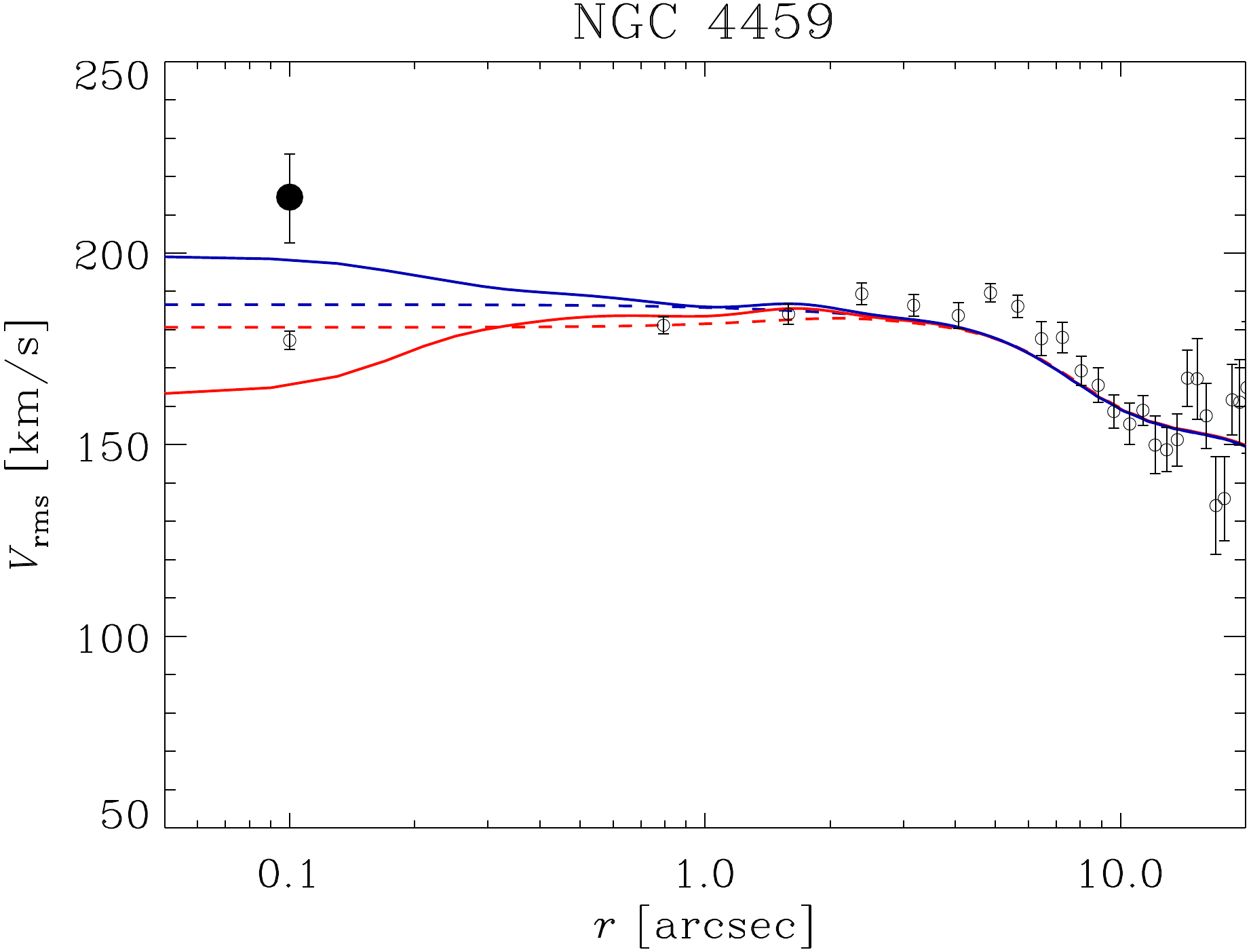}}
\caption{{\em Left-hand panels\/}: Comparison between the SAURON
  $V_{\rm rms}$ (black open circles) and second velocity moment from
  JAM best-fitting models with \mbh$=0$ \msun (red lines) and
  \mbh$=9.5 \cdot 10^{7}$ \msun\ (blue lines) along the major axis of
  NGC~4435. The dashed and continuous lines correspond to the JAM
  predictions for the SAURON and HST spatial resolution and sampling,
  respectively. The black filled circle corresponds to the
  sub-arcsecond \sigmas$_{\rm ,fix}$ measured from the G750M spectrum. The central
  SAURON $V_{\rm rms}$ and G750M \sigmas$_{\rm ,fix}$ are shifted to 0\farcs1 for
  comparison purposes only.  {\em Right-hand panels\/}: Same as
  above, but for NGC~4459 with \mbh$=0$ \msun (red lines) and
  \mbh$=1.1 \cdot 10^{8}$ \msun\ (blue lines).}
\label{fig:prediction}
\end{center}
\end{small}
\end{figure*}


Fig.~\ref{fig:prediction} further illustrates how the JAM models
  compare to the SAURON measurements by showing major-axis $V_{\rm
    rms}$ profiles for both data and models within a 0\farcs8-wide
  slit. Fig.~\ref{fig:prediction} shows both models with \mbh$ = 0$ \msun \ 
  and \mbh\ from the \msigmae\ relation, which allows to better 
  appreciate how these models differ by less than 10 \kms\ in 
  the central regions, as expected for models tailored to match 
  ground-based observations of limited spatial resolution. The large 
  discrepancy between data and models seen between about 1\arcsec\ 
  and 3\arcsec\ for NGC~4435 is due to the two-lobes structure of 
  the SAURON $V_{\rm rms}$ map, which is quite difficult to reproduce 
  (Fig.~\ref{fig:JAM}).

Having anchored the JAM models to match the SAURON data, we
  finally obtained the radial profiles of $V_{\rm rms}$ along the
  major axis of both galaxies by simply changing the spatial
  resolution and spatial sampling of the JAM models with \mbh$ = 0$
  \msun\ and \mbh\ from the \msigmae\ relation to match that of
  HST/STIS observations (Fig.~\ref{fig:prediction}).  As expected, the
  profiles at ground-based and HST resolution begin to differ only in
  the inner 1\arcsec, with a maximum difference of about 35 \kms\ in
  the centre in both the case of NGC~4435 and NGC~4459. This allowed a
  direct comparison with values of \sigmas$_{\rm ,fix}$ we measured from STIS
  spectra, which are also shown in Fig.~\ref{fig:prediction}.

For NGC~4435 the STIS \sigmas$_{\rm ,fix}=87.3^{+10.2}_{-8.8}$ \kms\ is
much smaller than the model prediction, at the HST resolution, of
$V_{\rm rms}(0) = 185.3 $ \kms\ for \mbh$=9.5 \cdot 10^{7}$ \msun. In
fact, our nuclear \sigmas value is even much smaller than the JAM
prediction for \mbh$ = 0$ \msun. On the other hand the STIS
\sigmas$_{\rm ,fix}=214.6^{+11.2}_{-12.0}$ \kms\ of NGC~4459 is almost consistent
within the 1$\sigma$ errors with the model prediction $V_{\rm rms}(0)
= 199.1 $ \kms\ for \mbh$=1.1 \cdot 10^{8}$ \msun.

Even though our reference JAM models are clearly somehow
  simplistic, in particular since they are based on the common
  assumption of constant vertical anisotropy profile and also in light
  of their limited ability to match the SAURON data in detail, we
  consider it unlikely that more sophisticated models for NGC~4435
  will be able to predict nuclear \sigmas\ values as low as those
  that we observe. For instance, to match the STIS \sigmas$_{\rm ,fix}$ alone we
  would currently need rather extreme vertical anisotropy $\beta_{z}$
  values of about $-0.9$. More mild $\beta_{z}$ variations, on the other
  hand, may help bridging the gap between the STIS measurements and
  the JAM model predictions in the case of NGC~4459.


\section{Conclusions}
\label{sec:conclusions}

We considered all the STIS spectra obtained with the G750M
  grating and bracketing the \Ha\ line available in the Hubble
  Data Archive and which were placed on the nucleus of a galaxy, in
  order to obtain stellar velocity dispersion measurements
  within sub-arcsecond apertures. In this way, we arrived at a sample
  of 28 relatively nearby galaxies ($D<70$ Mpc, see Table~\ref{tab:fit}),
  spanning a wide range of morphological types (from E to Scd),
  central nuclear activity (Seyfert 2, LINERs, \ion{H}{ii} nuclei) and
  nuclear stellar velocity dispersion measurements ($50 < $
  \sigmas\ $ < 270$ \kms). For these objects, we extracted the stellar
  kinematics by matching the nuclear STIS spectra using the single-age, 
  single-metallicity stellar population models of \citet{Vazdekis2010} 
  and both narrow and broad Gaussian emission lines. For 24 objects 
  we further constrained such a fit by using the optimal template 
  from a similar fit to the G430L low-resolution STIS spectra that were 
  also available for these galaxies, thus effectively fixing the stellar 
  population mix while matching the G750M spectra. The good agreement 
  between the nuclear \sigmas\ values obtained with and without using 
  such additional information, and additionally with the G750M
  \sigmas\ measurements of \citet{Batcheldor2013} in the \ion{Ca}{ii}
  triplet near-infrared wavelength region, gave us further confidence 
  in the \sigmas\ measurements for the 4 galaxies for which we could 
  only rely on the G750M spectra.

To start investigating the usefulness of such nuclear \sigmas\ 
measurements in constraining \mbh\ in nearby galaxies we then
considered the case of NGC~4435 and NGC~4459, since these indeed
appear to have quite different values of \mbh\ from ionised-gas
kinematical measurements despite showing similar ground-based
\sigmare\ values. Indeed, while NGC~4459 shows a \mbh\ value
consistent with the \msigmas\ relation \citep{Sarzi2001}, the \mbh\ of
NGC~4435 falls significantly below it \citep{Coccato2006}. Furthermore, 
these two galaxies feature quite different nuclear \sigmas\ values 
in our catalogue, with NGC~4435 displaying a particularly lower 
\sigmas\ value compared to NGC~4459 that could support the finding 
of \citet{Coccato2006}.

We looked into this puzzle by building Jeans axisymmetric stellar
  dynamical models for NGC~4435 and NGC~4459 starting from archival
  ACS images and while matching the available integral-field
  SAURON stellar kinematics. In addition, in these models we adopted
  either \mbh$= 0$ M$_\odot$ or the \mbh\ value from the
  \msigmae\ relation, which allowed us to predict the  
  $V_{\rm rms}(0) \simeq$ \sigmas\ by mimicking the HST spatial 
  resolution and sampling, while essentially holding to JAM the 
  values for stellar mass-to-light
  ratio, inclination, and anisotropy parameter that best matched the
  ground-based kinematics (see Section~\ref{sec:jam}).
In the case of NGC~4459 our nuclear \sigmas\ value
(\sigmas$_{\rm ,fix}=214.6^{+11.2}_{-12.0}$ \kms) was nearly consistent with the
high-spatial resolution prediction for $V_{\rm rms}(0)$ at the \mbh\ value
given by the \msigmae\ relation ($V_{\rm rms}(0)=199.1$ \kms), whereas
for NGC~4435 the nuclear \sigmas\ value (\sigmas$_{\rm ,fix}=87.3^{+10.2}_{-8.8}$
\kms) falls significantly below the model predictions even for \mbh$=
0$ M$_\odot$ ($V_{\rm rms}(0)=$ 151.6 \kms).

Although our models are somehow simplistic, these results lend further
support to the idea that the SBH of NGC~4435 is a genuine low outlier
in the \msigmae\ relation, which deserves more
investigation. For instance NGC~4435 would be an ideal target for
future near-infrared observations from either space (e.g. with JWST)
or from the ground while being assisted by adaptive optics (AO), such as
those presented by \citet{Krajnovic2018}, who indeed also found
evidence for undermassive SBH in two nearby early-type galaxies. 
AO assisted integral-field observations would also allow to establish 
if our nuclear \sigmas\ dip can be explained by the presence of a 
central dynamically cold component such as a nuclear disc, although 
we note that in that case the presence of a SBH would be betrayed by 
an increased central stellar rotation 
\citep[as also discussed by][]{Krajnovic2018}.
Theoretical works indicate that assessing the fraction of
undermassive black holes in nearby galaxies would be quite important
to constrain the processes driving the mass growth of SBHs and how
they settled in galactic nuclei \citep{Volonteri2011}.


\section*{Acknowledgements}

We acknowledge the anonymous referee. We are also grateful to John 
Magorrian, Bradley M. Peterson and S{\'e}bastien Viaene for their 
valuable comments. IP is supported by Fondazione Cassa di Risparmio 
di Padova e Rovigo (Cariparo) through the 2015 Ph.D. fellowship 
``Accurate mass determination of supermassive black holes in nearby
galaxies". EMC, EDB, LM, and AP acknowledge financial support
from Padua University through grants DOR1699945/16, DOR1715817/17,
DOR1885254/18, and BIRD164402/16. IP acknowledges the Centre for 
Astrophysics Research of the University of Hertfordshire, EMC and 
EDB acknowledge the Space Telescope Science Institute, and MS and BP
acknowledge the Institut d'Astrophysique de Paris for the hospitality 
while this paper was in progress. 
This research is based on observations made with the NASA/ESA Hubble 
Space Telescope, and obtained from the Hubble Legacy Archive 
(http://hla.stsci.edu/), which is a collaboration between the Space 
Telescope Science Institute (STScI/NASA), the Space Telescope European 
Coordinating Facility (ST-ECF/ESA) and the Canadian Astronomy Data 
Centre (CADC/NRC/CSA). This research made use of data from the Hubble 
Data Archive (https://archive.stsci.edu/hst/) and of the NASA/IPAC 
Extragalactic Database (http://ned.ipac.caltech.edu/).


\bibliographystyle{mnras}

\begin{thebibliography}{}
\makeatletter
\relax
\def\mn@urlcharsother{\let\do\@makeother \do\$\do\&\do\#\do\^\do\_\do\%\do\~}
\def\mn@doi{\begingroup\mn@urlcharsother \@ifnextchar [ {\mn@doi@}
  {\mn@doi@[]}}
\def\mn@doi@[#1]#2{\def\@tempa{#1}\ifx\@tempa\@empty \href
  {http://dx.doi.org/#2} {doi:#2}\else \href {http://dx.doi.org/#2} {#1}\fi
  \endgroup}
\def\mn@eprint#1#2{\mn@eprint@#1:#2::\@nil}
\def\mn@eprint@arXiv#1{\href {http://arxiv.org/abs/#1} {{\tt arXiv:#1}}}
\def\mn@eprint@dblp#1{\href {http://dblp.uni-trier.de/rec/bibtex/#1.xml}
  {dblp:#1}}
\def\mn@eprint@#1:#2:#3:#4\@nil{\def\@tempa {#1}\def\@tempb {#2}\def\@tempc
  {#3}\ifx \@tempc \@empty \let \@tempc \@tempb \let \@tempb \@tempa \fi \ifx
  \@tempb \@empty \def\@tempb {arXiv}\fi \@ifundefined
  {mn@eprint@\@tempb}{\@tempb:\@tempc}{\expandafter \expandafter \csname
  mn@eprint@\@tempb\endcsname \expandafter{\@tempc}}}

\bibitem[\protect\citeauthoryear{{Bacon} et~al.,}{{Bacon}
  et~al.}{2001}]{Bacon2001}
{Bacon} R.,  et~al., 2001, \mn@doi [\mnras] {10.1046/j.1365-8711.2001.04612.x},
  \href {http://adsabs.harvard.edu/abs/2001MNRAS.326...23B} {326, 23}

\bibitem[\protect\citeauthoryear{{Batcheldor}, {Axon}, {Valluri}, {Mandalou}
  \& {Merritt}}{{Batcheldor} et~al.}{2013}]{Batcheldor2013}
{Batcheldor} D.,  {Axon} D.,  {Valluri} M.,  {Mandalou} J.,   {Merritt} D.,
  2013, \mn@doi [\aj] {10.1088/0004-6256/146/3/67}, \href
  {http://adsabs.harvard.edu/abs/2013AJ....146...67B} {146, 67}

\bibitem[\protect\citeauthoryear{{Beifiori}, {Sarzi}, {Corsini}, {Dalla
  Bont{\`a}}, {Pizzella}, {Coccato}  \& {Bertola}}{{Beifiori}
  et~al.}{2009}]{Beifiori2009}
{Beifiori} A.,  {Sarzi} M.,  {Corsini} E.~M.,  {Dalla Bont{\`a}} E.,
  {Pizzella} A.,  {Coccato} L.,   {Bertola} F.,  2009, \mn@doi [\apj]
  {10.1088/0004-637X/692/1/856}, \href
  {http://adsabs.harvard.edu/abs/2009ApJ...692..856B} {692, 856}

\bibitem[\protect\citeauthoryear{{Beifiori}, {Courteau}, {Corsini}  \&
  {Zhu}}{{Beifiori} et~al.}{2012}]{Beifiori2012}
{Beifiori} A.,  {Courteau} S.,  {Corsini} E.~M.,   {Zhu} Y.,  2012, \mn@doi
  [\mnras] {10.1111/j.1365-2966.2011.19903.x}, \href
  {http://adsabs.harvard.edu/abs/2012MNRAS.419.2497B} {419, 2497}

\bibitem[\protect\citeauthoryear{{Binggeli}, {Sandage}  \&
  {Tammann}}{{Binggeli} et~al.}{1985}]{Binggeli1985}
{Binggeli} B.,  {Sandage} A.,   {Tammann} G.~A.,  1985, \mn@doi [\aj]
  {10.1086/113874}, \href {http://adsabs.harvard.edu/abs/1985AJ.....90.1681B}
  {90, 1681}

\bibitem[\protect\citeauthoryear{{Cappellari}}{{Cappellari}}{2002}]{Cappellari2002}
{Cappellari} M.,  2002, \mn@doi [\mnras] {10.1046/j.1365-8711.2002.05412.x},
  \href {http://adsabs.harvard.edu/abs/2002MNRAS.333..400C} {333, 400}

\bibitem[\protect\citeauthoryear{{Cappellari}}{{Cappellari}}{2008}]{Cappellari2008}
{Cappellari} M.,  2008, \mn@doi [\mnras] {10.1111/j.1365-2966.2008.13754.x},
  \href {http://adsabs.harvard.edu/abs/2008MNRAS.390...71C} {390, 71}

\bibitem[\protect\citeauthoryear{{Cappellari} \& {Emsellem}}{{Cappellari} \&
  {Emsellem}}{2004}]{Cappellari2004}
{Cappellari} M.,  {Emsellem} E.,  2004, \mn@doi [\pasp] {10.1086/381875}, \href
  {http://adsabs.harvard.edu/abs/2004PASP..116..138C} {116, 138}

\bibitem[\protect\citeauthoryear{{Cappellari} et~al.,}{{Cappellari}
  et~al.}{2007}]{Cappellari2007}
{Cappellari} M.,  et~al., 2007, \mn@doi [\mnras]
  {10.1111/j.1365-2966.2007.11963.x}, \href
  {http://adsabs.harvard.edu/abs/2007MNRAS.379..418C} {379, 418}

\bibitem[\protect\citeauthoryear{{Cappellari} et~al.,}{{Cappellari}
  et~al.}{2011}]{Cappellari2011}
{Cappellari} M.,  et~al., 2011, \mn@doi [\mnras]
  {10.1111/j.1365-2966.2010.18174.x}, \href
  {http://adsabs.harvard.edu/abs/2011MNRAS.413..813C} {413, 813}

\bibitem[\protect\citeauthoryear{{Cappellari} et~al.,}{{Cappellari}
  et~al.}{2013}]{Cappellari2013}
{Cappellari} M.,  et~al., 2013, \mn@doi [\mnras] {10.1093/mnras/stt562}, \href
  {http://adsabs.harvard.edu/abs/2013MNRAS.432.1709C} {432, 1709}

\bibitem[\protect\citeauthoryear{{Coccato}, {Sarzi}, {Pizzella}, {Corsini},
  {Dalla Bont{\`a}}  \& {Bertola}}{{Coccato} et~al.}{2006}]{Coccato2006}
{Coccato} L.,  {Sarzi} M.,  {Pizzella} A.,  {Corsini} E.~M.,  {Dalla Bont{\`a}}
  E.,   {Bertola} F.,  2006, \mn@doi [\mnras]
  {10.1111/j.1365-2966.2005.09901.x}, \href
  {http://adsabs.harvard.edu/abs/2006MNRAS.366.1050C} {366, 1050}

\bibitem[\protect\citeauthoryear{{Cortese}, {Bendo}, {Isaak}, {Davies}  \&
  {Kent}}{{Cortese} et~al.}{2010}]{Cortese2010}
{Cortese} L.,  {Bendo} G.~J.,  {Isaak} K.~G.,  {Davies} J.~I.,   {Kent} B.~R.,
  2010, \mn@doi [\mnras] {10.1111/j.1745-3933.2009.00808.x}, \href
  {http://adsabs.harvard.edu/abs/2010MNRAS.403L..26C} {403, L26}

\bibitem[\protect\citeauthoryear{{Cretton} \& {van den Bosch}}{{Cretton} \&
  {van den Bosch}}{1999}]{Cretton1999}
{Cretton} N.,  {van den Bosch} F.~C.,  1999, \mn@doi [\apj] {10.1086/306971},
  \href {http://adsabs.harvard.edu/abs/1999ApJ...514..704C} {514, 704}

\bibitem[\protect\citeauthoryear{{Dressel}, {Holfeltz}  \& {Quijano}}{{Dressel}
  et~al.}{2007}]{Dressel2007}
{Dressel} L.,  {Holfeltz} S.,   {Quijano} J.~K.,  2007, STIS Data Handbook,
  Version 5.0. STScI, Baltimore

\bibitem[\protect\citeauthoryear{{Emsellem}, {Monnet}  \& {Bacon}}{{Emsellem}
  et~al.}{1994}]{Emsellem1994}
{Emsellem} E.,  {Monnet} G.,   {Bacon} R.,  1994, \aap, \href
  {http://adsabs.harvard.edu/abs/1994A%26A...285..723E} {285, 723}

\bibitem[\protect\citeauthoryear{{Emsellem} et~al.,}{{Emsellem}
  et~al.}{2004}]{Emsellem2004}
{Emsellem} E.,  et~al., 2004, \mn@doi [\mnras]
  {10.1111/j.1365-2966.2004.07948.x}, \href
  {http://adsabs.harvard.edu/abs/2004MNRAS.352..721E} {352, 721}

\bibitem[\protect\citeauthoryear{{Falc{\'o}n-Barroso},
  {S{\'a}nchez-Bl{\'a}zquez}, {Vazdekis}, {Ricciardelli}, {Cardiel}, {Cenarro},
  {Gorgas}  \& {Peletier}}{{Falc{\'o}n-Barroso}
  et~al.}{2011}]{FalconBarroso2011}
{Falc{\'o}n-Barroso} J.,  {S{\'a}nchez-Bl{\'a}zquez} P.,  {Vazdekis} A.,
  {Ricciardelli} E.,  {Cardiel} N.,  {Cenarro} A.~J.,  {Gorgas} J.,
  {Peletier} R.~F.,  2011, \mn@doi [\aap] {10.1051/0004-6361/201116842}, \href
  {http://adsabs.harvard.edu/abs/2011A%26A...532A..95F} {532, A95}

\bibitem[\protect\citeauthoryear{{Falc{\'o}n-Barroso}
  et~al.,}{{Falc{\'o}n-Barroso} et~al.}{2017}]{FalconBarroso2017}
{Falc{\'o}n-Barroso} J.,  et~al., 2017, \mn@doi [\aap]
  {10.1051/0004-6361/201628625}, \href
  {http://adsabs.harvard.edu/abs/2017A%26A...597A..48F} {597, A48}

\bibitem[\protect\citeauthoryear{{Ferrarese} et~al.,}{{Ferrarese}
  et~al.}{2006}]{Ferrarese2006}
{Ferrarese} L.,  et~al., 2006, \mn@doi [\apjs] {10.1086/501350}, \href
  {http://adsabs.harvard.edu/abs/2006ApJS..164..334F} {164, 334}

\bibitem[\protect\citeauthoryear{{Gonzaga} et~al.,}{{Gonzaga}
  et~al.}{2012}]{DrizzlePac2012}
{Gonzaga} S.,  et~al., 2012, The DrizzlePac Handbook, Version 1.0. STScI,
  Baltimore

\bibitem[\protect\citeauthoryear{{Graham}}{{Graham}}{2016}]{Graham2016c}
{Graham} A.~W.,  2016, in {Laurikainen} E.,  {Peletier} R.,   {Gadotti} D.,
  eds,  Astrophysics and Space Science Library Vol. 418, Galactic Bulges.
  p.~263

\bibitem[\protect\citeauthoryear{{G{\"u}ltekin}, {Richstone}, {Gebhardt},
  {Faber}, {Lauer}, {Bender}, {Kormendy}  \& {Pinkney}}{{G{\"u}ltekin}
  et~al.}{2011}]{Gultekin2011}
{G{\"u}ltekin} K.,  {Richstone} D.~O.,  {Gebhardt} K.,  {Faber} S.~M.,  {Lauer}
  T.~R.,  {Bender} R.,  {Kormendy} J.,   {Pinkney} J.,  2011, \mn@doi [\apj]
  {10.1088/0004-637X/741/1/38}, \href
  {http://adsabs.harvard.edu/abs/2011ApJ...741...38G} {741, 38}

\bibitem[\protect\citeauthoryear{{Ho}, {Filippenko}  \& {Sargent}}{{Ho}
  et~al.}{1997}]{Ho1997}
{Ho} L.~C.,  {Filippenko} A.~V.,   {Sargent} W.~L.~W.,  1997, \mn@doi [\apjs]
  {10.1086/313041}, \href {http://adsabs.harvard.edu/abs/1997ApJS..112..315H}
  {112, 315}

\bibitem[\protect\citeauthoryear{{Kenney}, {Tal}, {Crowl}, {Feldmeier}  \&
  {Jacoby}}{{Kenney} et~al.}{2008}]{Kenney2008}
{Kenney} J.~D.~P.,  {Tal} T.,  {Crowl} H.~H.,  {Feldmeier} J.,   {Jacoby}
  G.~H.,  2008, \mn@doi [\apjl] {10.1086/593300}, \href
  {http://adsabs.harvard.edu/abs/2008ApJ...687L..69K} {687, L69}

\bibitem[\protect\citeauthoryear{{Kormendy} \& {Ho}}{{Kormendy} \&
  {Ho}}{2013}]{Kormendy2013}
{Kormendy} J.,  {Ho} L.~C.,  2013, \mn@doi [\araa]
  {10.1146/annurev-astro-082708-101811}, \href
  {http://adsabs.harvard.edu/abs/2013ARA%26A..51..511K} {51, 511}

\bibitem[\protect\citeauthoryear{{Kormendy} et~al.,}{{Kormendy}
  et~al.}{1997}]{Kormendy1997}
{Kormendy} J.,  et~al., 1997, \mn@doi [\apjl] {10.1086/310720}, \href
  {http://adsabs.harvard.edu/abs/1997ApJ...482L.139K} {482, L139}

\bibitem[\protect\citeauthoryear{{Kormendy}, {Fisher}, {Cornell}  \&
  {Bender}}{{Kormendy} et~al.}{2009}]{Kormendy2009}
{Kormendy} J.,  {Fisher} D.~B.,  {Cornell} M.~E.,   {Bender} R.,  2009, \mn@doi
  [\apjs] {10.1088/0067-0049/182/1/216}, \href
  {http://adsabs.harvard.edu/abs/2009ApJS..182..216K} {182, 216}

\bibitem[\protect\citeauthoryear{{Krajnovi{\'c}} \& {Jaffe}}{{Krajnovi{\'c}} \&
  {Jaffe}}{2004}]{Krajnovic2004}
{Krajnovi{\'c}} D.,  {Jaffe} W.,  2004, \mn@doi [\aap]
  {10.1051/0004-6361:20040359}, \href
  {http://adsabs.harvard.edu/abs/2004A%26A...428..877K} {428, 877}

\bibitem[\protect\citeauthoryear{{Krajnovi{\'c}} et~al.,}{{Krajnovi{\'c}}
  et~al.}{2011}]{Krajnovic2011}
{Krajnovi{\'c}} D.,  et~al., 2011, \mn@doi [\mnras]
  {10.1111/j.1365-2966.2011.18560.x}, \href
  {http://adsabs.harvard.edu/abs/2011MNRAS.414.2923K} {414, 2923}

\bibitem[\protect\citeauthoryear{{Krajnovi{\'c}} et~al.,}{{Krajnovi{\'c}}
  et~al.}{2018}]{Krajnovic2018}
{Krajnovi{\'c}} D.,  et~al., 2018, \mn@doi [\mnras] {10.1093/mnras/sty778},
  \href {http://adsabs.harvard.edu/abs/2018MNRAS.tmp..752K} {}

\bibitem[\protect\citeauthoryear{{Krist}, {Hook}  \& {Stoehr}}{{Krist}
  et~al.}{2011}]{Krist2011}
{Krist} J.~E.,  {Hook} R.~N.,   {Stoehr} F.,  2011, in Optical Modeling and
  Performance Predictions V. p. 81270J

\bibitem[\protect\citeauthoryear{{Kuntschner} et~al.,}{{Kuntschner}
  et~al.}{2010}]{Kuntschner2010}
{Kuntschner} H.,  et~al., 2010, \mn@doi [\mnras]
  {10.1111/j.1365-2966.2010.17161.x}, \href
  {http://adsabs.harvard.edu/abs/2010MNRAS.408...97K} {408, 97}

\bibitem[\protect\citeauthoryear{{Lablanche} et~al.,}{{Lablanche}
  et~al.}{2012}]{Lablanche2012}
{Lablanche} P.-Y.,  et~al., 2012, \mn@doi [\mnras]
  {10.1111/j.1365-2966.2012.21343.x}, \href
  {http://adsabs.harvard.edu/abs/2012MNRAS.424.1495L} {424, 1495}

\bibitem[\protect\citeauthoryear{{Lucas} et~al.,}{{Lucas}
  et~al.}{2016}]{Lucas2016}
{Lucas} R.~A.,  et~al., 2016, ACS Instrument Handbook, Version 8.0. STScI,
  Baltimore

\bibitem[\protect\citeauthoryear{{McDermid} et~al.,}{{McDermid}
  et~al.}{2015}]{McDermid2015}
{McDermid} R.~M.,  et~al., 2015, \mn@doi [\mnras] {10.1093/mnras/stv105}, \href
  {http://adsabs.harvard.edu/abs/2015MNRAS.448.3484M} {448, 3484}

\bibitem[\protect\citeauthoryear{{Nelder} \& {Mead}}{{Nelder} \&
  {Mead}}{1965}]{Nelder1965}
{Nelder} J.~A.,  {Mead} R.,  1965, Computer Journal, 7, 308

\bibitem[\protect\citeauthoryear{{Nowak}, {Saglia}, {Thomas}, {Bender},
  {Davies}  \& {Gebhardt}}{{Nowak} et~al.}{2008}]{Nowak2008}
{Nowak} N.,  {Saglia} R.~P.,  {Thomas} J.,  {Bender} R.,  {Davies} R.~I.,
  {Gebhardt} K.,  2008, \mn@doi [\mnras] {10.1111/j.1365-2966.2008.13960.x},
  \href {http://adsabs.harvard.edu/abs/2008MNRAS.391.1629N} {391, 1629}

\bibitem[\protect\citeauthoryear{{Pagotto} et~al.,}{{Pagotto}
  et~al.}{2017}]{Pagotto2017}
{Pagotto} I.,  et~al., 2017, \mn@doi [Astronomische Nachrichten]
  {10.1002/asna.201713370}, \href
  {http://adsabs.harvard.edu/abs/2017AN....338..841P} {338, 841}

\bibitem[\protect\citeauthoryear{{Pavlovsky} et~al.,}{{Pavlovsky}
  et~al.}{2004}]{Pavlovsky2004}
{Pavlovsky} C.,  et~al., 2004, ACS Instrument Handbook, Version 5.0. STScI,
  Baltimore

\bibitem[\protect\citeauthoryear{{Riley} et~al.,}{{Riley}
  et~al.}{2017}]{Riley2017}
{Riley} A.,  et~al., 2017, {STIS Instrument Handbook, Version 16.0. STScI,
  Baltimore}

\bibitem[\protect\citeauthoryear{{Saglia} et~al.,}{{Saglia}
  et~al.}{2016}]{Saglia2016}
{Saglia} R.~P.,  et~al., 2016, \mn@doi [\apj] {10.3847/0004-637X/818/1/47},
  \href {http://adsabs.harvard.edu/abs/2016ApJ...818...47S} {818, 47}

\bibitem[\protect\citeauthoryear{{S{\'a}nchez-Bl{\'a}zquez}
  et~al.,}{{S{\'a}nchez-Bl{\'a}zquez} et~al.}{2006}]{SanchezBlazquez2006}
{S{\'a}nchez-Bl{\'a}zquez} P.,  et~al., 2006, \mn@doi [\mnras]
  {10.1111/j.1365-2966.2006.10699.x}, \href
  {http://adsabs.harvard.edu/abs/2006MNRAS.371..703S} {371, 703}

\bibitem[\protect\citeauthoryear{{Sani}, {Marconi}, {Hunt}  \&
  {Risaliti}}{{Sani} et~al.}{2011}]{Sani2011}
{Sani} E.,  {Marconi} A.,  {Hunt} L.~K.,   {Risaliti} G.,  2011, \mn@doi
  [\mnras] {10.1111/j.1365-2966.2011.18229.x}, \href
  {http://adsabs.harvard.edu/abs/2011MNRAS.413.1479S} {413, 1479}

\bibitem[\protect\citeauthoryear{{Sarzi}, {Rix}, {Shields}, {Rudnick}, {Ho},
  {McIntosh}, {Filippenko}  \& {Sargent}}{{Sarzi} et~al.}{2001}]{Sarzi2001}
{Sarzi} M.,  {Rix} H.-W.,  {Shields} J.~C.,  {Rudnick} G.,  {Ho} L.~C.,
  {McIntosh} D.~H.,  {Filippenko} A.~V.,   {Sargent} W.~L.~W.,  2001, \mn@doi
  [\apj] {10.1086/319724}, \href
  {http://adsabs.harvard.edu/abs/2001ApJ...550...65S} {550, 65}

\bibitem[\protect\citeauthoryear{{Sarzi} et~al.,}{{Sarzi}
  et~al.}{2002}]{Sarzi2002}
{Sarzi} M.,  et~al., 2002, \mn@doi [\apj] {10.1086/338351}, \href
  {http://adsabs.harvard.edu/abs/2002ApJ...567..237S} {567, 237}

\bibitem[\protect\citeauthoryear{{Sarzi}, {Rix}, {Shields}, {Ho}, {Barth},
  {Rudnick}, {Filippenko}  \& {Sargent}}{{Sarzi} et~al.}{2005}]{Sarzi2005}
{Sarzi} M.,  {Rix} H.-W.,  {Shields} J.~C.,  {Ho} L.~C.,  {Barth} A.~J.,
  {Rudnick} G.,  {Filippenko} A.~V.,   {Sargent} W.~L.~W.,  2005, \mn@doi
  [\apj] {10.1086/428637}, \href
  {http://adsabs.harvard.edu/abs/2005ApJ...628..169S} {628, 169}

\bibitem[\protect\citeauthoryear{{Sarzi} et~al.,}{{Sarzi}
  et~al.}{2006}]{Sarzi2006}
{Sarzi} M.,  et~al., 2006, \mn@doi [\mnras] {10.1111/j.1365-2966.2005.09839.x},
  \href {http://adsabs.harvard.edu/abs/2006MNRAS.366.1151S} {366, 1151}

\bibitem[\protect\citeauthoryear{{Savorgnan} \& {Graham}}{{Savorgnan} \&
  {Graham}}{2015}]{SavorgnanGraham2015}
{Savorgnan} G.~A.~D.,  {Graham} A.~W.,  2015, \mn@doi [\mnras]
  {10.1093/mnras/stu2259}, \href
  {http://adsabs.harvard.edu/abs/2015MNRAS.446.2330S} {446, 2330}

\bibitem[\protect\citeauthoryear{{Savorgnan} \& {Graham}}{{Savorgnan} \&
  {Graham}}{2016}]{SavorgnanGraham2016}
{Savorgnan} G.~A.~D.,  {Graham} A.~W.,  2016, \mn@doi [\apjs]
  {10.3847/0067-0049/222/1/10}, \href
  {http://adsabs.harvard.edu/abs/2016ApJS..222...10S} {222, 10}

\bibitem[\protect\citeauthoryear{{Schlafly} \& {Finkbeiner}}{{Schlafly} \&
  {Finkbeiner}}{2011}]{Schlafly2011}
{Schlafly} E.~F.,  {Finkbeiner} D.~P.,  2011, \mn@doi [\apj]
  {10.1088/0004-637X/737/2/103}, \href
  {http://adsabs.harvard.edu/abs/2011ApJ...737..103S} {737, 103}

\bibitem[\protect\citeauthoryear{{Scott} et~al.,}{{Scott}
  et~al.}{2013}]{Scott2013}
{Scott} N.,  et~al., 2013, \mn@doi [\mnras] {10.1093/mnras/sts422}, \href
  {http://adsabs.harvard.edu/abs/2013MNRAS.432.1894S} {432, 1894}

\bibitem[\protect\citeauthoryear{{Shankar} et~al.,}{{Shankar}
  et~al.}{2016}]{Shankar2016}
{Shankar} F.,  et~al., 2016, \mn@doi [\mnras] {10.1093/mnras/stw678}, \href
  {http://adsabs.harvard.edu/abs/2016MNRAS.460.3119S} {460, 3119}

\bibitem[\protect\citeauthoryear{{Vazdekis}, {S{\'a}nchez-Bl{\'a}zquez},
  {Falc{\'o}n-Barroso}, {Cenarro}, {Beasley}, {Cardiel}, {Gorgas}  \&
  {Peletier}}{{Vazdekis} et~al.}{2010}]{Vazdekis2010}
{Vazdekis} A.,  {S{\'a}nchez-Bl{\'a}zquez} P.,  {Falc{\'o}n-Barroso} J.,
  {Cenarro} A.~J.,  {Beasley} M.~A.,  {Cardiel} N.,  {Gorgas} J.,   {Peletier}
  R.~F.,  2010, \mn@doi [\mnras] {10.1111/j.1365-2966.2010.16407.x}, \href
  {http://adsabs.harvard.edu/abs/2010MNRAS.404.1639V} {404, 1639}

\bibitem[\protect\citeauthoryear{{Verdoes Kleijn}, {van der Marel}, {de Zeeuw},
  {Noel-Storr}  \& {Baum}}{{Verdoes Kleijn} et~al.}{2002}]{VerdoesKleijn2002}
{Verdoes Kleijn} G.~A.,  {van der Marel} R.~P.,  {de Zeeuw} P.~T.,
  {Noel-Storr} J.,   {Baum} S.~A.,  2002, \mn@doi [\aj] {10.1086/344073}, \href
  {http://adsabs.harvard.edu/abs/2002AJ....124.2524V} {124, 2524}

\bibitem[\protect\citeauthoryear{{Vika}, {Driver}, {Cameron}, {Kelvin}  \&
  {Robotham}}{{Vika} et~al.}{2012}]{Vika2012}
{Vika} M.,  {Driver} S.~P.,  {Cameron} E.,  {Kelvin} L.,   {Robotham} A.,
  2012, \mn@doi [\mnras] {10.1111/j.1365-2966.2011.19881.x}, \href
  {http://adsabs.harvard.edu/abs/2012MNRAS.419.2264V} {419, 2264}

\bibitem[\protect\citeauthoryear{{Vollmer}, {Braine}, {Combes}  \&
  {Sofue}}{{Vollmer} et~al.}{2005}]{Vollmer2005}
{Vollmer} B.,  {Braine} J.,  {Combes} F.,   {Sofue} Y.,  2005, \mn@doi [\aap]
  {10.1051/0004-6361:20041389}, \href
  {http://adsabs.harvard.edu/abs/2005A%26A...441..473V} {441, 473}

\bibitem[\protect\citeauthoryear{{Volonteri}, {Natarajan}  \&
  {G{\"u}ltekin}}{{Volonteri} et~al.}{2011}]{Volonteri2011}
{Volonteri} M.,  {Natarajan} P.,   {G{\"u}ltekin} K.,  2011, \mn@doi [\apj]
  {10.1088/0004-637X/737/2/50}, \href
  {http://adsabs.harvard.edu/abs/2011ApJ...737...50V} {737, 50}

\bibitem[\protect\citeauthoryear{{Walsh}, {van den Bosch}, {Gebhardt},
  {Yildirim}, {G{\"u}ltekin}, {Husemann}  \& {Richstone}}{{Walsh}
  et~al.}{2015}]{Walsh2015}
{Walsh} J.~L.,  {van den Bosch} R.~C.~E.,  {Gebhardt} K.,  {Yildirim} A.,
  {G{\"u}ltekin} K.,  {Husemann} B.,   {Richstone} D.~O.,  2015, \mn@doi [\apj]
  {10.1088/0004-637X/808/2/183}, \href
  {http://adsabs.harvard.edu/abs/2015ApJ...808..183W} {808, 183}

\bibitem[\protect\citeauthoryear{{Walsh}, {van den Bosch}, {Gebhardt},
  {Y{\i}ld{\i}r{\i}m}, {G{\"u}ltekin}, {Husemann}  \& {Richstone}}{{Walsh}
  et~al.}{2017}]{Walsh2017}
{Walsh} J.~L.,  {van den Bosch} R.~C.~E.,  {Gebhardt} K.,  {Y{\i}ld{\i}r{\i}m}
  A.,  {G{\"u}ltekin} K.,  {Husemann} B.,   {Richstone} D.~O.,  2017, \mn@doi
  [\apj] {10.3847/1538-4357/835/2/208}, \href
  {http://adsabs.harvard.edu/abs/2017ApJ...835..208W} {835, 208}

\bibitem[\protect\citeauthoryear{{de Vaucouleurs}, {de Vaucouleurs}, {Corwin},
  {Buta}, {Paturel}  \& {Fouqu\'e}}{{de Vaucouleurs}
  et~al.}{1991}]{deVaucouleurs1991}
{de Vaucouleurs} G.,  {de Vaucouleurs} A.,  {Corwin} Jr. H.~G.,  {Buta} R.~J.,
  {Paturel} G.,   {Fouqu\'e} P.,  1991, Third Reference Catalogue of Bright
  Galaxies, Springer-Verlag, New York (RC3)

\bibitem[\protect\citeauthoryear{{van Dokkum}}{{van Dokkum}}{2001}]{Dokkum2001}
{van Dokkum} P.~G.,  2001, \mn@doi [\pasp] {10.1086/323894}, \href
  {http://adsabs.harvard.edu/abs/2001PASP..113.1420V} {113, 1420}

\bibitem[\protect\citeauthoryear{{van den Bosch}}{{van den
  Bosch}}{2016}]{vandenBosch2016}
{van den Bosch} R.~C.~E.,  2016, \mn@doi [\apj] {10.3847/0004-637X/831/2/134},
  \href {http://adsabs.harvard.edu/abs/2016ApJ...831..134V} {831, 134}

\bibitem[\protect\citeauthoryear{{van den Bosch}, {Gebhardt}, {G{\"u}ltekin},
  {van de Ven}, {van der Wel}  \& {Walsh}}{{van den Bosch}
  et~al.}{2012}]{vandenBosch2012}
{van den Bosch} R.~C.~E.,  {Gebhardt} K.,  {G{\"u}ltekin} K.,  {van de Ven} G.,
   {van der Wel} A.,   {Walsh} J.~L.,  2012, \mn@doi [\nat]
  {10.1038/nature11592}, \href
  {http://adsabs.harvard.edu/abs/2012Natur.491..729V} {491, 729}

\bibitem[\protect\citeauthoryear{{van den Bosch}, {Gebhardt}, {G{\"u}ltekin},
  {Y{\i}ld{\i}r{\i}m}  \& {Walsh}}{{van den Bosch}
  et~al.}{2015}]{vandenBosch2015}
{van den Bosch} R.~C.~E.,  {Gebhardt} K.,  {G{\"u}ltekin} K.,
  {Y{\i}ld{\i}r{\i}m} A.,   {Walsh} J.~L.,  2015, \mn@doi [\apjs]
  {10.1088/0067-0049/218/1/10}, \href
  {http://adsabs.harvard.edu/abs/2015ApJS..218...10V} {218, 10}

\makeatother
\end{thebibliography}


\bsp	
\label{lastpage}
\end{document}